**Geometrical joke(r?)s for SETI.**

**R. T. Faizullin**

**OmSTU, Omsk, Russia**

Since the beginning of radio era long delayed echoes (LDE) were traced. They are the most likely candidates for extraterrestrial communication, the so-called "paradox of Stormer" or "world echo". By LDE we mean a radio signal with a very long delay time and abnormally low energy loss. Unlike the well-known echoes of the delay in 1/7 seconds, the mechanism of which have long been resolved, the delay of radio signals in a second, ten seconds or even minutes is one of the most ancient and intriguing mysteries of physics of the ionosphere. Nowadays it is difficult to imagine that at the beginning of the century any registered echo signal was treated as extraterrestrial communication:

“Notable changes occurred at a fixed time and the analogy among the changes and numbers was so clear, that I could not provide any plausible explanation. I'm familiar with natural electrical interference caused by the activity of the Sun, northern lights and telluric currents, but I was sure, as it is possible to be sure in anything, that the interference was not caused by any of common reason. Only after a while it came to me, that the observed interference may occur as the result of conscious activities. I'm overwhelmed by the the feeling, that I may be the first men to hear greetings transmitted from one planet to the other... Despite the signal being weak and unclear it made me certain that soon people, as one, will direct their eyes full of hope and affection towards the sky, overwhelmed by good news: People! We got the message from an unknown and distant planet. And it sounded: one...two....three...” Nikola Tesla, 1900

It was a different story with LDE. The idea that radio-echo may be an artificial phenomenon, something like an original card of extraterrestrial satellite, which attracts our attention was put forward only after the publication of a brief note by the astronomer Ronald Bracewell in the Nature journal in 1960. In the beginning, LDE was taken as evidence of the existence of specific fast-moving clouds of plasma in cosmic space that can not only simply reflect radio signals, like the Earth's ionosphere, but it can also focus the original signal so that the power of the reflected signal exceeds a third of the power source! The turning point was the letter from the engineer Jorgen Halse to a famous astrophysicist Carl Stormer.

“...At the end of the summer of 1927 I repeatedly heard signals from the Dutch short-wave transmitting station PCJJ at Eindhoven. At the same time as I heard these I also heard echoes. I heard the usual echo which goes round the Earth with an interval of about 1/7th of a second as well as a weaker echo about three seconds after the principal echo had gone. When the principal signal was especially strong, I suppose the amplitude for the last echo three seconds later, lay between 1/10 and 1/20 of the principal signal in strength. From where this echo comes I cannot say for the present, I can only confirm that I really heard it." [2]

To confirm these facts the astronomer Stormer, the physicist Van der Pol (the famous equation of Van der Pol) and the engineer Hulse organized a series of experiments that were designed to verify the presence of the phenomenon and its frequency of exposure.

In 1927, a transmitter, located in Eindhoven, began to transmit pulses that were recorded by Halse in Oslo. Originally, each signal was a series of three points of Morse. These signals were repeated every 5 seconds. In September, the regime of the transmitter was changed. The intervals were increased to 20 seconds. The experiment was not described in all details, because the publication of the experimental conditions were presented at the conference proceedings with strict limitations. On October 11th, 1928, a series of radio-echo were finally recorded. Van der Pol mentioned it in his telegram to Stormer and Halse: "Last night, our signals were accompanied by an echo, the echo time ranged between 3 and 15 seconds, half of the echoes lasted more than 8 seconds! " Halse and Stormer in turn, confirmed the acquisition of these echoes in Oslo, several seriess of echo were obtained. Recorded echo delay time ranged from 3 seconds to 3.5 minutes! In November 1929 an experiment was completed.

The following 5 series of echo delays were registered [3]:

1) 15,9,4,8,13,8,12,10,9,5,8,7,6
2) 12,14,14,12,8
3) 12,5,8
4) 12,8,5,14,14,15,12,7,5,5,13,8,8,8,13,9,10,7,14,6,9,5,9
5) 8,11,15,8,13,3,8,8,8,12,15,13,8,8 (seconds)

In May 1929 J. Galle and G. Talon made new successful research of LDE phenomenon:

"... In May 1929, a French expedition was in Indo-China to study an eclipse of the Sun. J B Galle and G Talon, captain of the naval vessel L'Inconstant, had orders to study the effects of the eclipse on radio propagation, particularly long delayed echoes. They used a 500 watt transmitter with a 20 meter aerial attached to an 8 meter mast, powered by the generators of the Indo-China Hydrographic Service vessel La Perouse. The two aptly named ships sailed from Saigon on May 2nd, and on May 5th they conducted test transmissions in "la baie de Penitencier", PouloCondere, and detected long delayed echoes. Weather conditions prevented work on May 6th and 7th, but on the 8th the ships were back on station and transmitted for the first ten minutes of every half hour. On May 9th, the day of the eclipse, signals were sent for nearly six hours with one 20 minute break, and again for ten minutes in every half hour the following day. Two dots were sent every 30 seconds on 25 metres wavelength, varying in a fixed musical sequences to aid correct identification and timing of the echoes. Large numbers of echoes were heard, clearly divided into two groups: weak echoes, about 1/100 the original signal strength, and strong ones 1/3 to 1/5 the intensity of the transmitted signal, with no significant relation between strength and delay time. (These intensities are too great for natural reflection at such apparent distances, but no-one seems to have thought of that at the time.) In their preliminary report Galle and Talon said echoes stopped altogether during the totality of the eclipse, but in fact they paused 3 1/2 minutes before the eclipse became total and began again half way through it. Delay times ranged from 1 second to 30 seconds, though two 31 second echoes and of 32 seconds were heard between 15.40 and 16.00 on the day of the eclipse. 1 and 2 second echoes might seem impossible for a probe in the Moon's orbit, but for an extraordinary circumstance. At 14h 19m 29s on the day of the eclipse the operator "forgot" to send the required dots, but 5 and 10 second echoes were heard nonetheless. From this Galle and Talon concluded that some echoes might have 40 seconds delay or more: either their musical tone sequences let them down, or they were unable to believe evidence that the probe was anticipating their signals as it transmitted its "replies" [4, 5, 6].

In 1934 LDE phenomenon was observed by E. Appleton. His data presented as the histogram is one of the most precise materials on LDE experiments. [1] (Fig 1 ).

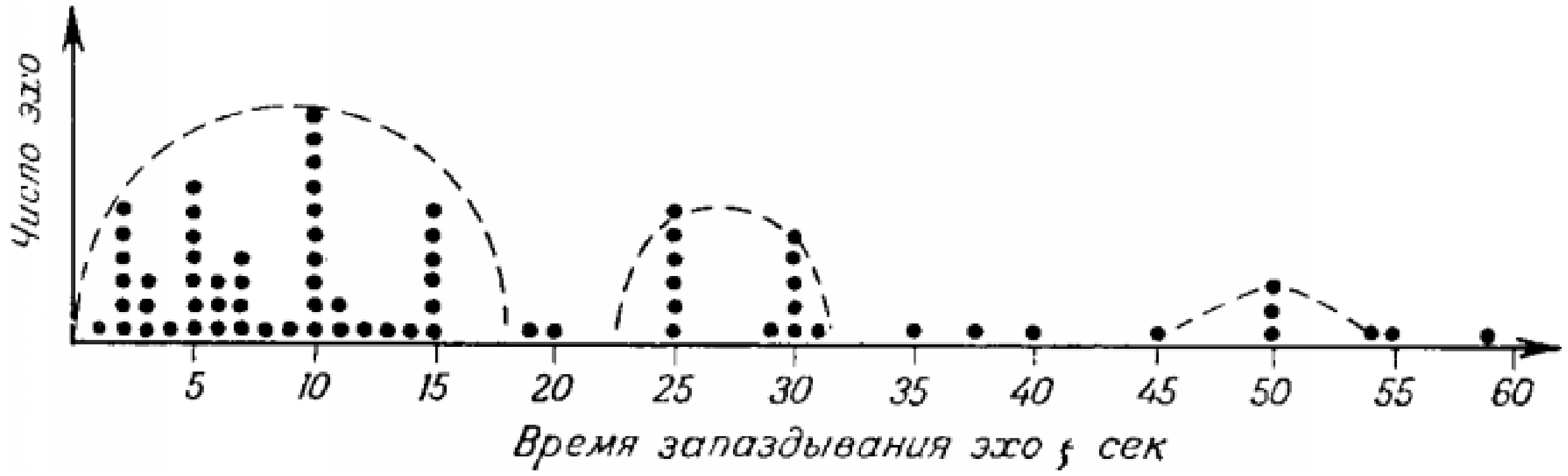

Fig. 1. Number delays and time delays (sec). For more convenience pictures descriptions on the Russian and on the English too.

In 1967 the experiments to trace LDE were preformed at Stanford University by F. Crawford. He was able to confirm the phenomenon, but especially long echo delays and series that were detected in 1920-30's was not found. Delays with 2 and 8 seconds were the most frequent with the rate shift and time contraction compared to the time between the impulses of the original signal. The research of the LDE data provides an interesting observation – in any new wave band, which is just beginning to be used, this phenomenon is very clear and in series, just like in 1920's, but after a while the echo starts to fade and it is no longer periodical.

Lunen, an English astronomer, noticed that LDE observed in 1920's lacked time compression and Doppler frequency shift. Also the rate of Stormer frequency remained the same, despite the time delay. It is rather difficult to explain when we take into consideration only natural signals. Natural echo signals with 3 seconds to 3 minutes delay time can not possibly be of the same rate, because of the dispersion of the signal, as the wave created by a transmitter is not a coherent laser pulse!

Duncan Lunen was the first to hypothesize that the echo of Stormer series of numbers may be a signal of interstellar probe. He also stated that changes in delay time may be an attempt to deliver some information. Lunen considered that this information is about a planet, where the probe was sent from. Using star maps, he concluded that the probe was originally from Epsilon Bootis:

" ... If the data points are plotted with delay time on the y-axis (normal scientific practice, followed by all the 1920s experimenter' s who presented their results graphically), nothing significant appears. With delay time on the x-axis, however, the graph looks more like an intelligent signal . There is a vertical "barrier" at 8 seconds dividing the diagram into two parts of an equal area; on the left there is a single dot, at three seconds, which was unique in being an exact repeat of the transmitted signal, three dots, the other echoes being 2 second long dashes. On the right of the barrier the main figure has a striking but incomplete resemblance to the constellation Bootes, the Herdsman . If the 3 second dot is transplanted across the barrier to a corresponding position on the right, it occupies the position of the star Epsilon Bootis and so completes the constellation figure. ..."

Lunen studied one of the Stormer series of 1928:

8, 11, 15, 8, 13, 3, 8, 8, 8, 12, 15, 13, 8, 8.

Arbitrariness of Lunen's geometrical plottings was proven almost at once and not even by sceptics, but by enthusiasts. Using different method of deciphering, amateur astronomers from Bulgaria found out that the probe came from the star called Zeta Leo. Shpilevsky method of deciphering showed that the home planet of the prob is Tau Ceti.

This situation is very similar to one described in the novel His Master's Voice by Stanislaw Lem: a brief note that appeared in press and contained some hints proving the extraterrestrial contact. This note drowned in the sea of pseudoscientific publications which make any sensible person doubt the reliability of the provided information. There was no need for secret services to interfere or for the disinformation to be provided in case of Lunen's graph. It can be viewed as a procedure of verification made by enthusiasts, which was mentioned above. The fact that such graphs can be easily done is shown on fig. 3.

This figure shows the coordinates of impulses registered in META. Each of this impulses equaled a well-known signal “Wow!” and was detected on the “hot” line – a wave of 21cm long! So what do we see when we link the coordinates in the order determined by the dates, we finally get an outline of some space ship, which is just like a plane, can signal us with its lights from time to time. (fig. 3)

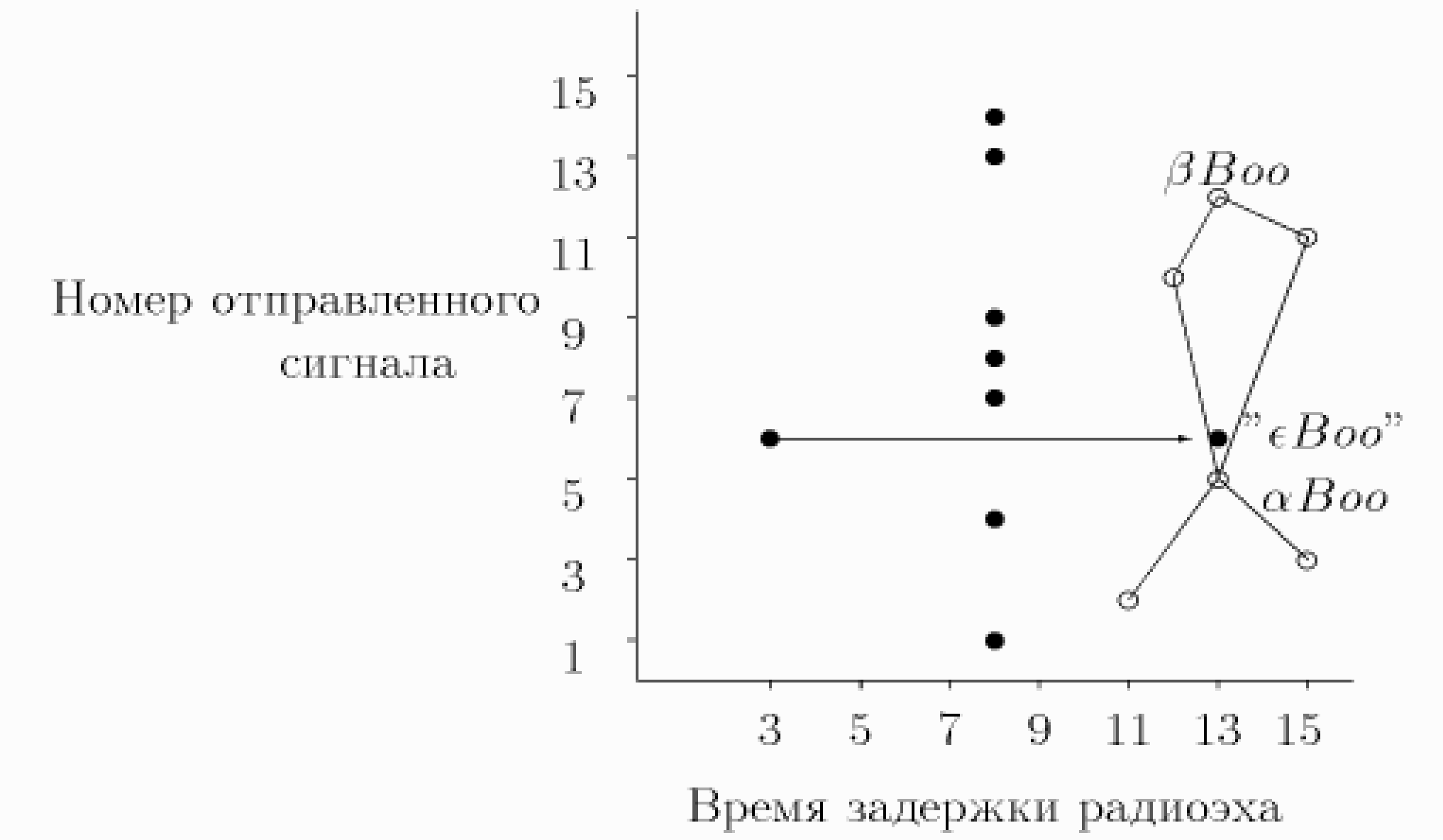

График Лунена. Если расположить радиоэхо в соответствие с их временем задержки, то семь точек вытянутся в вертикальный барьер, правее которого шесть точек образуют конфигурацию напоминающую абрис созвездия Волопаса. Левее лежит только одна точка. Если ее зеркально отобразить относительно барьера, то ее местоположение будет "почти" совпадать с местоположением эпсилон Волопаса.

Fig. 2. If we place radio echoes according to their delay time, then 7 points will stretch in a horizontal barrier. 6 points on the right form a pattern that looks like Bootes. There is just one point on the left. Its reflectional symmetry about the barrier almost matches the position of Epsilon Bootis.

It seemed that this is it! Unfortunately it was only an artefact, the device that was used to scan the sky was able to cover only a small vertical interval. Day after day this interval was rising up and after it reached the peak it started descending. The next figure shows the way signals are arranged on the wave double the size of the previous graph – 42 cm. The figure demonstrates that the outline happened purely by chance.(fig.4)

So what was Lunen's mistake? We can always find a configuration in the starry sky that may match the given random configuration of points. This is just one of the cases of Ramsey's theorem. If there are enough objects or points, we can always find specific configuration (figure) with special properties. It won't be ideal of course, but the more points or objects we have, the more exact the copy will be.

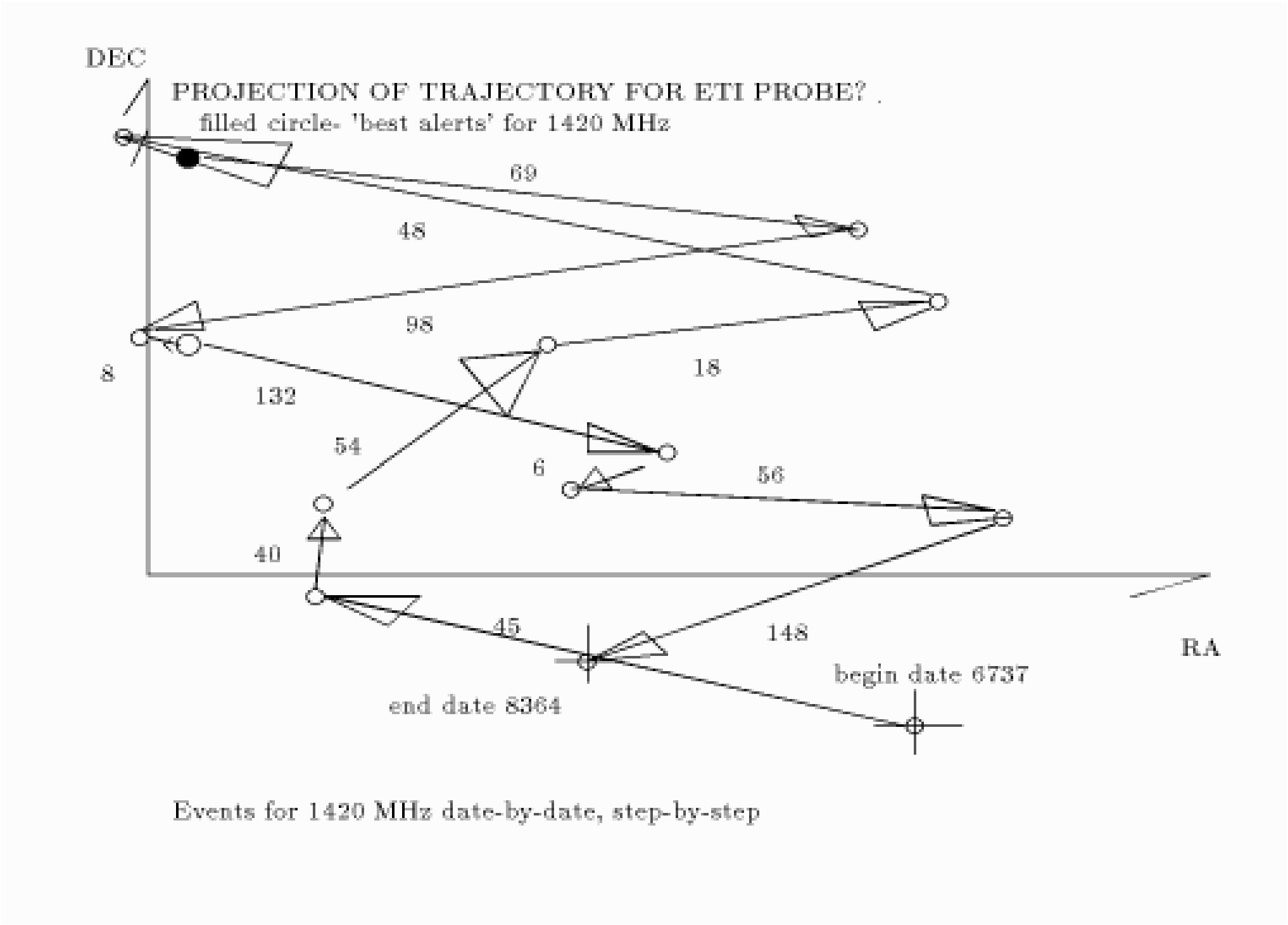

Fig. 3.

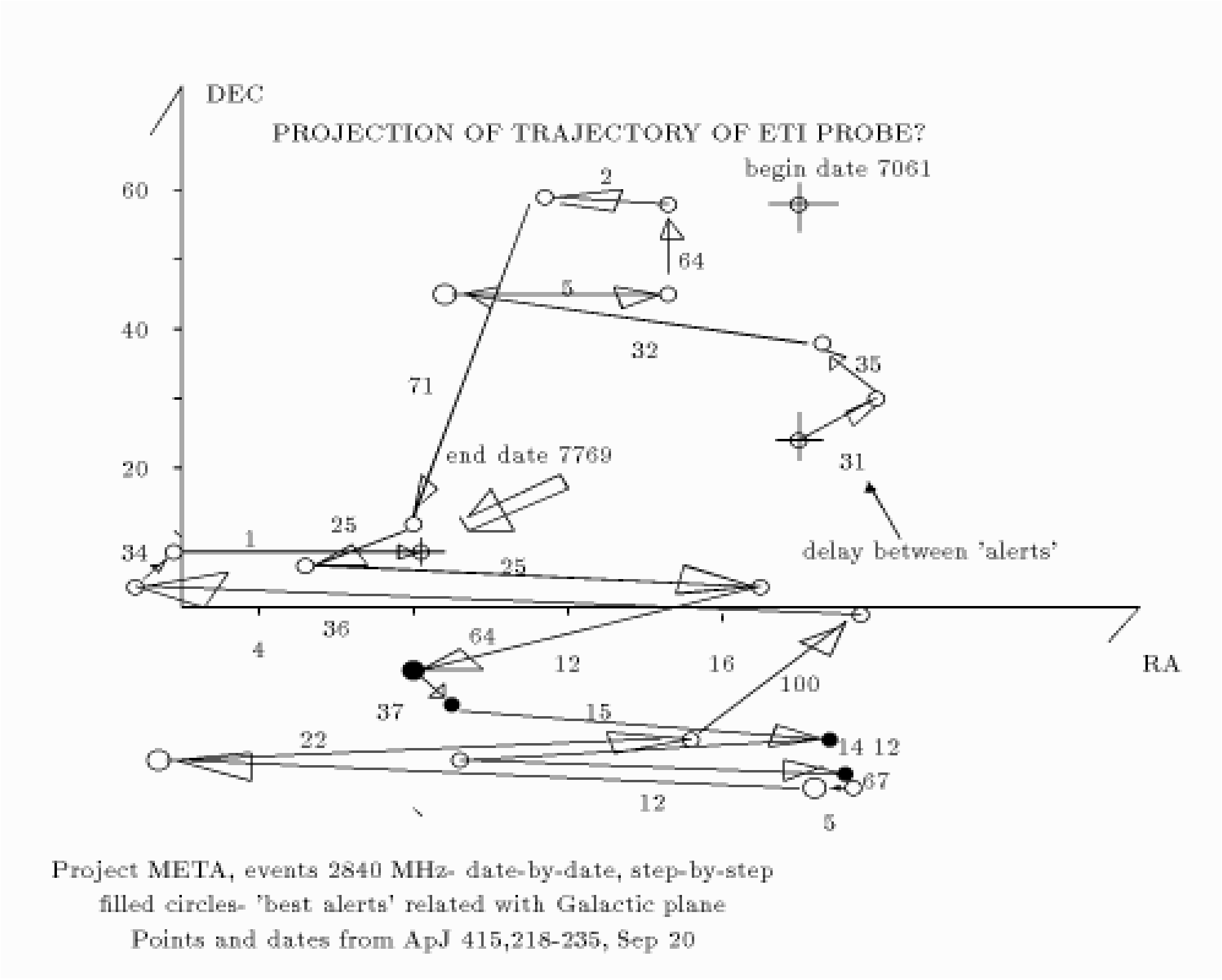

Fig. 4.

Considering delay time, Lunen tried to outline a figure, that would look like constellation. There are many constellations, bat they are more cultural artifacts, which means that the way of grouping the stars was random. It was just a phantom of artistic imagination of Babel and Greek cultures not more.

Before moving on to the discussion of hypothesis sated in this article, it is necessary to consider some limitations and regulations of a contact message.

It should be noted that the transmitting procedure as well as its content represent some well thought out standard which leads to success. Therefore the message itself should be regarded as a product of the highest technology which was tested dozens of times and most likely was not even developed by the civilization that sends this message at this time. It should be underlined, that we do not talk only about technology in its usual sense, but also about social technologies.

So what conclusions can we make? First of all, the guaranty of successful identification of the artificiality of the message should be expected. This means that if we record the message, we can decipher it with high probability. So, the message should naturally be recorded first and then documented in the annals of science, religion and other long-term organizations. Simultaneous interpretation of the message, or it is better to say understanding, should not be expected at once, it is highly doubtable.

Should the message be redundant? It is a complicated question. What exactly is meant by redundancy and whether it is useful? Consider an example related to a famous signal "Wow!". Suppose, that from the same point of the celestial sphere, where the signal was originally received, analogical signal is transmitted, so the time intervals between the signals correspond as prime numbers (the so-called call sign). Will this be the evidence of artificiality? Beyond all doubts. But, we should also consider the receiver of the message. Will the receiver be able to scan this particular part of the sky for years or even for centuries? It is not even the matter of scientific interest, but of financial opportunities.

Will the space of the recipient be free from satellite radio signals? Besides, who can insure, that there are no physical processes in nature, caused by spontaneous charges, that occur in intervals similar to prime numbers? Thus, the redundancy of the "signal" may be the basis for determining its natural origin.

(Recent research of spontaneous scintillations in the interstellar clouds of hydrogen casts doubt on the possibility of the existence of such channel of communication, because the natural scintillation create noise. This noise can be very complicated. Due to the complexity it can be confused with artificial signal. It's also a case of Ramsey's theory.)

Instead of the redundancy an independence should be expected. We gradually moved on to discuss the possible content of a signal or as it is commonly called - call signs. Examples of call signs already exist, although they are examples of terrestrial origin (Pioneers 10, 11 and the subsequent, long distance Voyager probes). Based on the real messages, you can reasonably make assumptions about the initial (historically speaking) form of standard message - a little bit about yourself, the coordinates and something about Space.

Will anything change then?

"When a man discovers what moves the stars, then the Sphinx would laugh, and the human race will fail". (inscription on the walls of Abu Simbel)

No matter how much time passes and how different the levels of development of the civilization are, the contact message is always addressed to someone very much interested in stars. This should be something that separates men from beast, even from a very clever beast.

One more important thing worth mentioning, again leads us to the stars. It is a time interval during which the message will be of any interest to the receiver. In our case this interval should be equal the period of human existence. It's about 100 000 years!

It can be declared that we know all the basic facts about stars, but it's not new...

All the above mentioned allows us to propose the following hypothesis and to formulate requirements for its verification:

1. Delay times of the signal should be interpreted as numbers of celestial bodies in some objective arrangement.
2. We should seek for unexpected geometric properties, applied to figures, outlined on the stellar field. In case of success, we should try to determine mathematical and physical sense of this properties.
3. The obtained data should be confirmed for several independent series of LDEs, which were detected in different times and by different researchers to eliminate subjective interpretations and falsification of data.
4. The results of the experiments should go through standard academic approbation (conference speech, scientific publications (with no references to SETI problem)). Apart form hypothesis of Contact, it should prove the importance of this mathematical and physical facts.

Before moving on to the demonstration of concrete results, lets study all the points in details. Point 1 describes what we mean by celestial bodies. Planets? Not all of their parts can be observed by the naked eye, so the requirement of independence, which is to ensure that the knowledge would be accessible to the maximum number of observers, fails in case of planets. The nearest stars to us? Just like the nearest planets, they can not be seen without telescope, because they are so small in mass and luminosity. Besides the speed of movement of small stars is so high, that their position on the stellar field has changed in thousands of years and presumed figure " fades". This is also true about galaxies, bright nebulas and objects with low luminosity. So that leaves us with the brightest stars. Their speed is low, they can do easily observed. People always were interested in them.

What kind of criteria should be used to sort them? Should we sort them according to distance, mass or luminosity?

Stellar parallaxes are not quite exact even nowadays. The masses are secondary values, so it is better to use luminosity or, to be more exact, visual stellar magnitude (V)

Visual magnitudes can be determined with high accuracy using photoelectric method. At the same time the magnitudes can be calculated with the help of expert judgment on the earliest times of the development of civilization.

The table below provides a list of 50 most luminous stars (arranged according to V)

(Kulikovsky [8], 1985)

| | | | l | b |
|---|---|---|---|---|
| 1. Sirius | 6h 43m | -16 35 | 227 | -9 |
| 2. Canopus | 6h 23m | -52 40 | 261 | -25 |
| 3. Arcturus | 14h 13m | +19 27 | 15 | +69 |
| 4. Vega | 18h 35m | +38 44 | 67 | +19 |
| 5. Tolimán | 14h 36m | -60 38 | 316 | +1 |
| 6. Capella | 5h 13m | +45 57 | 163 | +5 |
| 7. Rigel | 5h 12m | -08 15 | 209 | -25 |
| 8. Procyon | 7h 37m | +05 21 | 219 | +13 |
| 9. Betelgeuse | 5 52 | +07 24 | 200 | -9 |
| 10. Achernar | 1 36 | -57 29 | 291 | -59 |
| 11. Hadar | 14 00 | -60 08 | 314 | .0 |
| 12. Altair | 19 48 | +08 44 | 48 | -9 |
| 13. Alpha Crucis | 12 24 | -62 49 | 300 | 0 |
| 14 Aldebaran | 4 33 | +16 25 | 181 | -20 |
| 15. Antares | 16 26 | -26 19 | 352 | +15 |
| 16. Spica | 13 23 | -10 54 | 316 | +51 |
| 17. Pollux | 7 42 | +28 09 | 192 | +23 |
| 18. Fomalhaut | 22 55 | -29 53 | 20 | -65 |
| 19. Mimosa | 12 45 | -59 09 | 302 | +3 |
| 20. Deneb | 20 40 | +45 06 | 84 | +2 |
| 21. Regulus | 10 06 | +12 13 | | |
| 22. Adar | 6 57 | -28 54 | | |
| 23. Castor | 7 31 | +32 00 | | |
| 24. Saul | 17 30 | -37 04 | | |
| 25. Bellatrix | 5 22 | +06 18 | | |
| 26. Gamma Crucis | 12 28 | -56 50 | | |
| 27. El Nath | 5 23 | +28 34 | | |
| 28. Miaplacidus | 9 13 | -69 31 | | |
| 29. Alnilam | 5 34 | -01 14 | | |
| 30. Al Nair | 22 05 | -47 12 | | |
| 31. Aliot | 12 52 | +56 14 | | |
| 32. Alnitak | 5 38 | -01 58 | | |
| 33. Dubhe | 11 01 | +62 01 | | |
| 34. Alpha Persei | 3 21 | +49 41 | | |
| 35. Gamma Steam | 8 08 | -47 12 | | |
| 36. Kaus Aust | 18 21 | -34 25 | | |
| 37. Weesen | 7 06 | -26 19 | | |
| 38. Benetnash | 13 46 | +49 34 | | |
| 39. Avior | 8 22 | -59 21 | | |
| 40. Theta Cap | 17 34 | -42 58 | | |
| 41. Menkalinan | 5 56 | +44 57 | | |
| 42. Alchen | 6 35 | +16 27 | | |
| 43. Metalen | 16 43 | -68 56 | | |
| 44. Delta Steam | 8 43 | -54 31 | | |

| | | |
|---|---|---|
| 45. Peacock | 20 22 | -56 54 |
| 46. Mirza | 6 20 | -17 56 |
| 47. Alfard | 9 25 | -08 26 |
| 48. Mira | 2 17 | -03 12 |
| 49. Hamal | 2 04 | +23 14 |
| 50. Polar | 1 49 | +89 02 |

(Aristillus was the first to arrange the stars according to the visual magnitude. He asked several people with exceptionally acute eye-sight to estimate the level of brightness of stars. Hipparchus got his scale by averaging out the results. This was the first scientific application of the so-called Delphi method. It should be noted that Hipparchus scale is different from the one made with the help of the photoelectric method).

Point 2 determines what kind of figures we should search for. Lunen's failure demonstrates that we should search not for anthropomorphic symbols but mathematic figures, that are not random and reflect the laws of stellar dynamics. But how does stellar magnitude correspond to the dynamics? It is known that full stellar luminosity (L) is a function of it's mass. To be more specific, it is proportional to the mass raised in the 3$^{rd}$ power plus some small quantity. Visual stellar magnitude is proportional to $L/R^2$. This means that stellar magnitude is an automodel characteristics, which can characterise both mass and distance.

The 3$^{rd}$ point is nothing more then a usual requirement to any experiment, roughly saying, it's when repeatability excludes chance.

The same is true about the 4$^{th}$ point. Let's consider that by chance we got some strange outline of a figure which resembles mathematical figure, but the fact of the publication itself serves as a prof of the importance of the result. Publications involve not only the description of the phenomenon, but also some sort of development, such as confirmation of proposed physical model by numerical experiments, creation of new algorithms, using data of the observation of this phenomenon, etc. On the other hand, is it possible to get scientific facts using random numbers?

We turn to the immediate consideration of the facts and hypotheses.

Consider the so-called first Stormer's series:

15, 9, 4, 8, 13, 8, 12, 10, 9, 5, 8, 7, 6.

It is naturally expected that this series, if it's an artificial message, contains a "call sign"- some geometrical fact, illustration, similar to that offered by Gauss (Pythagorean theorem, illustrated with deforestation in Siberia or channels of burning oil in Sahara).

So what is so special about these stars? If we study them using common Cartesian coordinate system in space or on stellar field, there's nothing unusual, but lets use spherical coordinate system.

How can we build it? Lets examine stellar sphere of radius one. Imagine that we as observers are on the Earth and we are surrounded by a huge sphere. The stars are located outside the sphere and they are projected on the surface of the sphere. Neglecting their distance, we can only see bright spots (points) on the sphere. Each point lies in the intersection of line with the sphere, the line is drawn through the center of the sphere and the real location of the star. Then the star can have its angular coordinates (l, b).

The plane passes through the center of the sphere. If we choose some ray that passes the center of the sphere, we can calculate the angle in relation to the ray. This is coordinate *l.* The arc is an angle *b* between the plane and the ray directed from the center of the sphere on the point of this sphere that doesn't lie on this plane.

It is obvious the angle *l* lies in the range between 0 and 360 degrees. *b* can vary between -90 and +90 degrees. Now, if we look at the system of coordinates where OL substitutes axes OX and axes OY is replaced by OB, then we get some amount of stars within the rectangle [0,360]x[-90,90]. It may seem that the difference can only be in more comfortable representation of data, but it is not that simple. For example, instead of the plane that goes through the center of the sphere, we can choose some random plane or the plane of rotation of the system of the observer. Thus we choose dynamically dedicated coordinate system with the moment of rotation. *l* and *b* are regarded as galactic coordinate system, where the plane of section of celestial sphere is the same as the plane of Galaxy, *l*=0 was chosen randomly of course, but it turned out to be not exactly a random choice, because *l* == 0 corresponds to the direction to the center of the Galaxy.

The other system of coordinates correlates to the plane of ecliptic system and corresponds to the plane of rotation and to the moment of rotation of Solar system. The third spherical system of coordinates is equatorial, where the plane of section is perpendicular to the Earth rotational axis.

It should be noted that because of the random choice of *l*=0 and periodicity the geometrical figure we get is not a rectangle but a cylinder. The erected from *l* = 0 and *l* = 360 set up the same lines.

It may seem that given coordinate system is far-fetched and uncomfortable, but in reality Descartes' idea is even more uncomfortable. Spherical coordinate systems allow us to map the stellar sphere on the limited part of plane. This systems are natural, because of local spherical symmetry and isotropy of space and natural connection of *l* coordinate with the moment of rotation of the system.

Then in the galactic coordinate system we get an amazing figure which can easily fulfil the function of call-sign. (fig 5.)

We have 8 lines, which can be transformed into 2 tripods of parallel lines and a different pair of parallel lines. (fig 6., fig 7).

It should be pointed out that the selection from the numbers, that match the first series of Stormer, doesn't provide this kind of symmetry, and, moreover, the probability of an accidental but such accurate illustration of the property of parallelism when we randomly trace the given points equals the number of order 1 / (12!). (Code, if it's a code, can easily be deciphered. When combining the first 5 points in the order of 15, 9, 4, 8, 13, we get two parallel lines and then it is quite clear that there should be drawn lines 13, 8, 12, to obtain intervals (8, 12), (12, 10), etc. It is probably worth remembering the idea of very simple codes to decipher and existing error correcting codes).

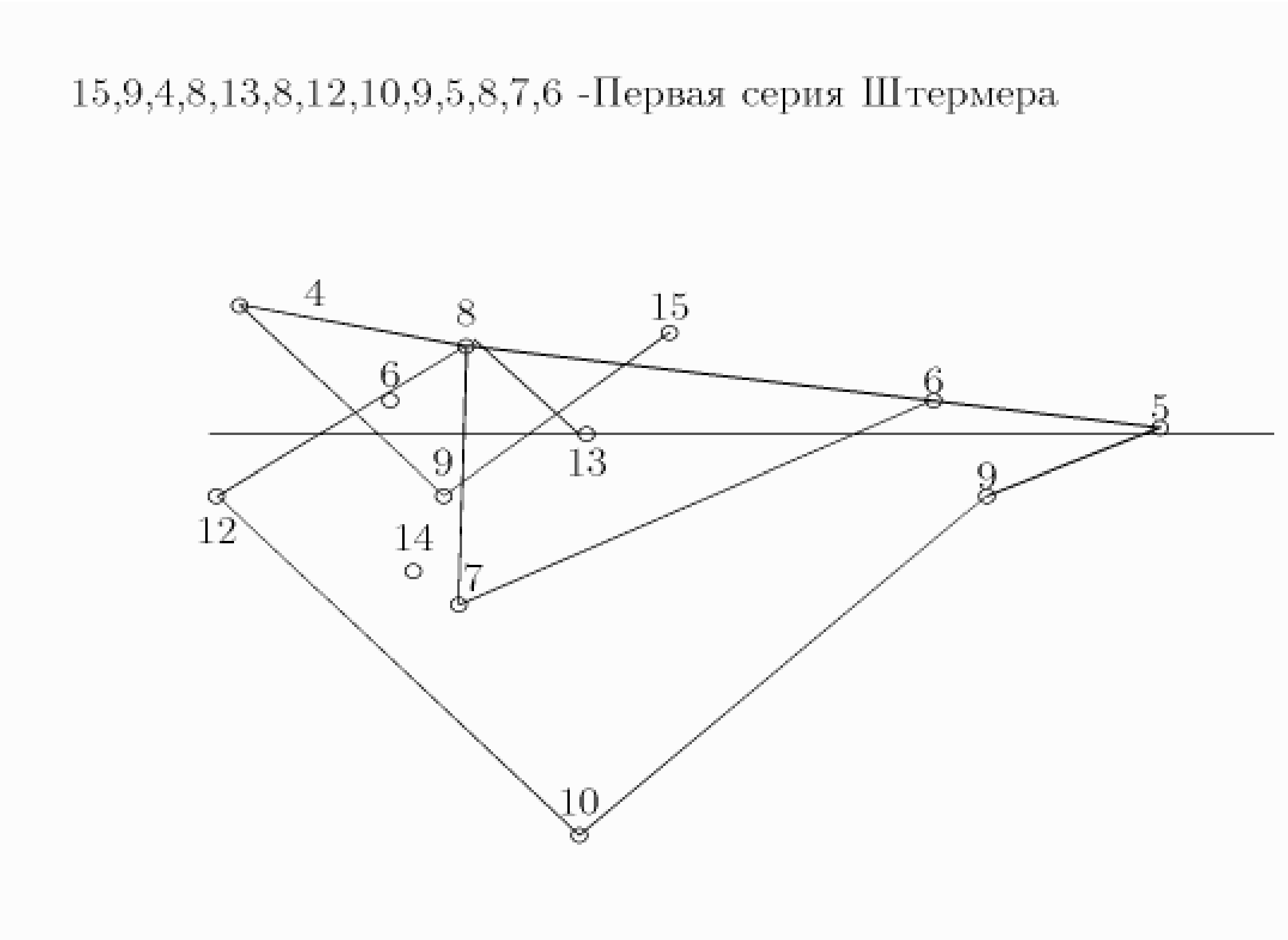

Fig. 5.

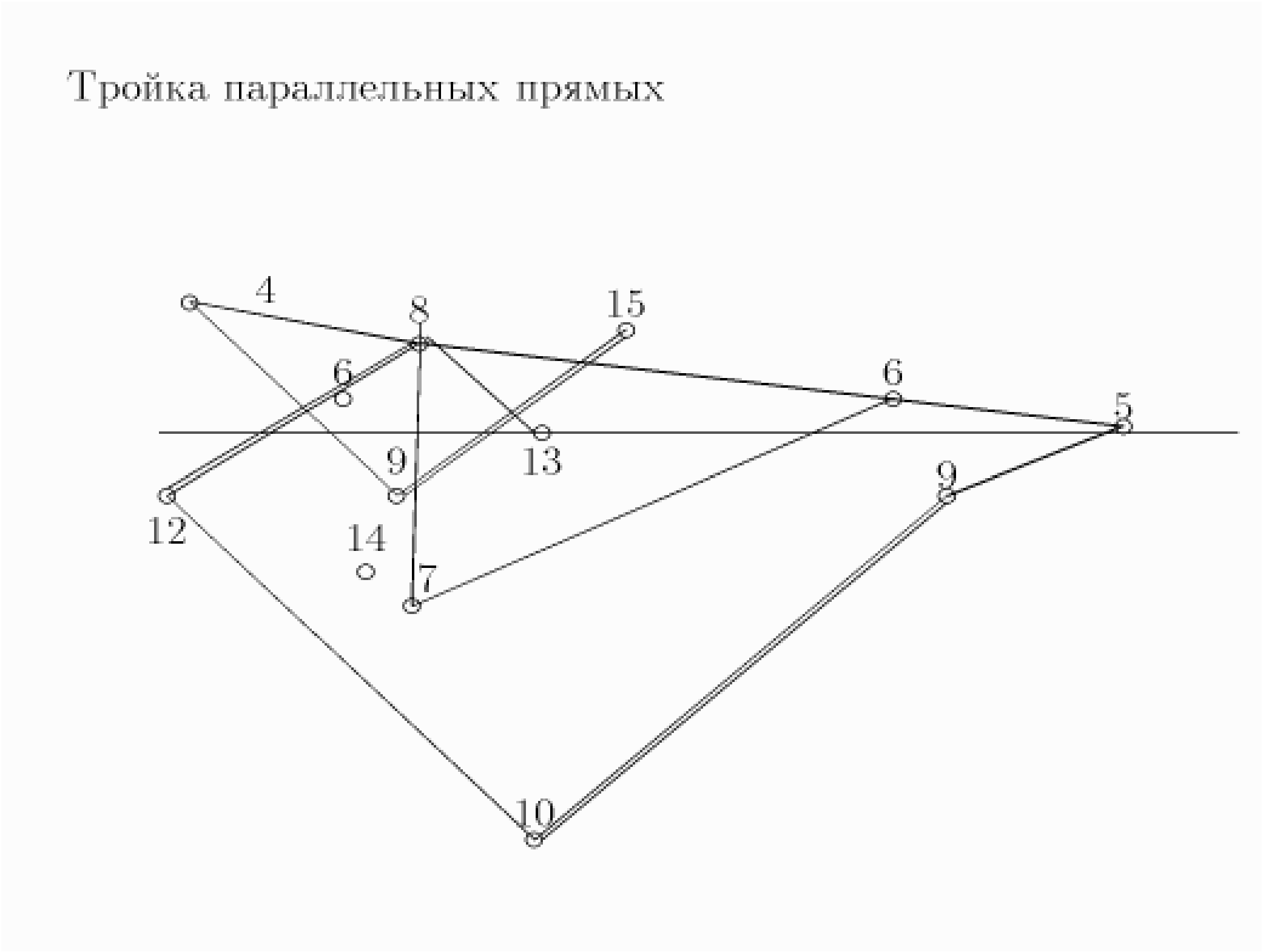

Fig. 6.

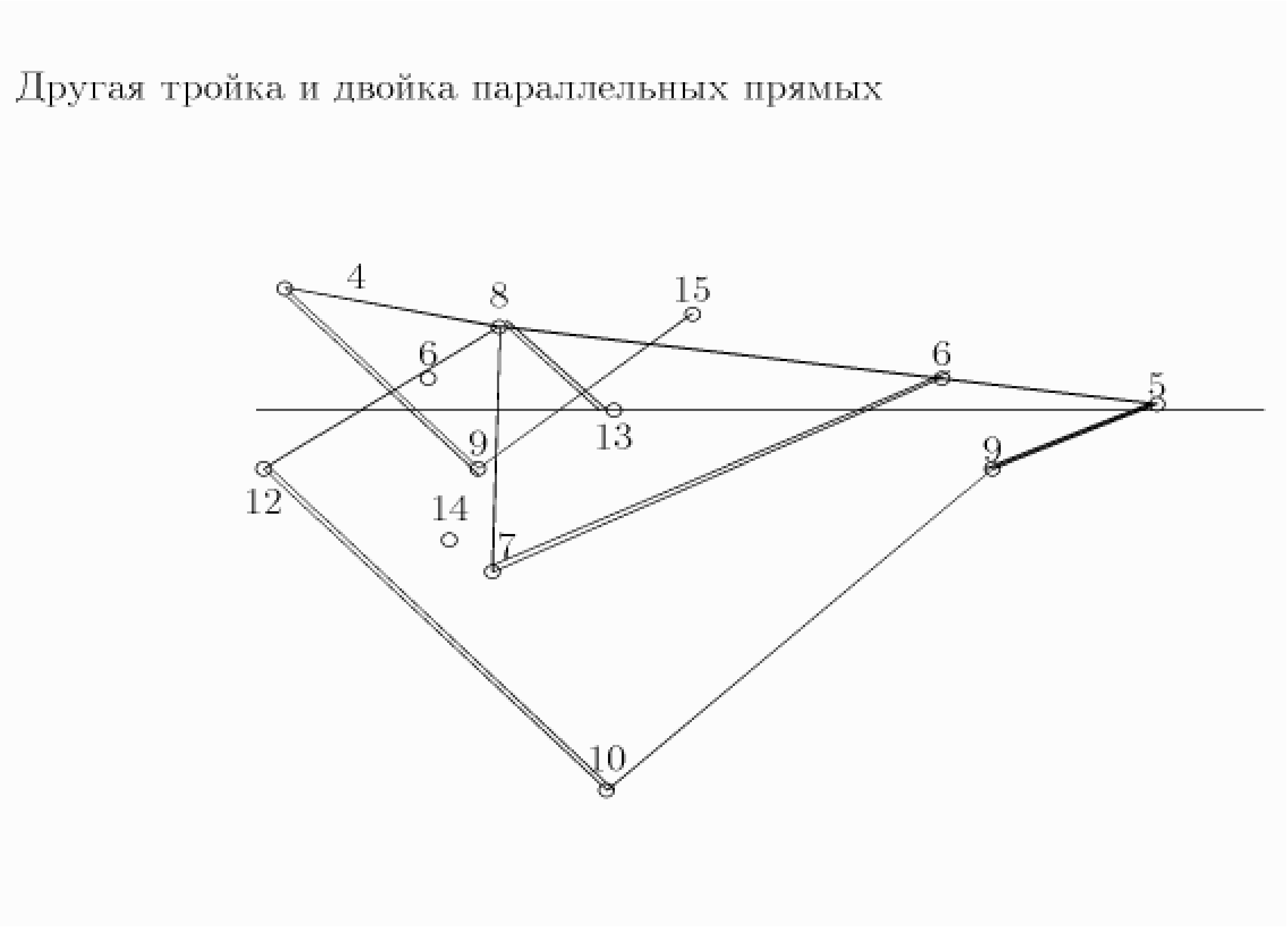

Fig. 7.

The question now arises of whether we should only restrict ourselves to the list of the brightest stars? If they are considered as means of attracting of attention, it can work out well. On the other hand how can we reach “inner assurance” of the contractor of the Contact, that it's a real Message, not a game of chance?

Lets consider some other lists of astrophysical objects and their natural arrangement. For example, lets study the list of the closest stars to the Sun, which was made by Kulikovsky in his work. (fig. 35):

| | | | |
|---|---|---|---|
| 1 | $\alpha$ Cen | | |
| 2 | Barnard's “runaway” star | | |
| 3 | Wolf 359 | 244 | 55 |
| 4 | +36 | 185.5 | 64.75 |
| 5 | $\alpha$ Cma | 229 | -9 |
| 6 | L726-8 | -176 | 75.5 |
| 7 | Ross 154 | 12.5 | -9.3 |
| 8 | Ross 248 | 110 | -17 |
| 9 | $\varepsilon$ Eri | 195.6 | -48 |
| 11 | Ross 128 | 270 | 59 |
| 10 | L 789-6 | 44.5 | -56 |
| 13 | $\varepsilon$ Ind | 336.3 | -48 |
| 12 | 61 Cyg | 82.5 | -5.6 |
| 14 | $\alpha$ Cmi | 219 | 13 |
| 15 | +59 | 90.5 | 24.5 |
| 16 | +43 | 116 | -20 |

The closest stellar systems (according to their arrangement in 1985, P.G. Kulikovsky). There is again a wide range of parallel lines.

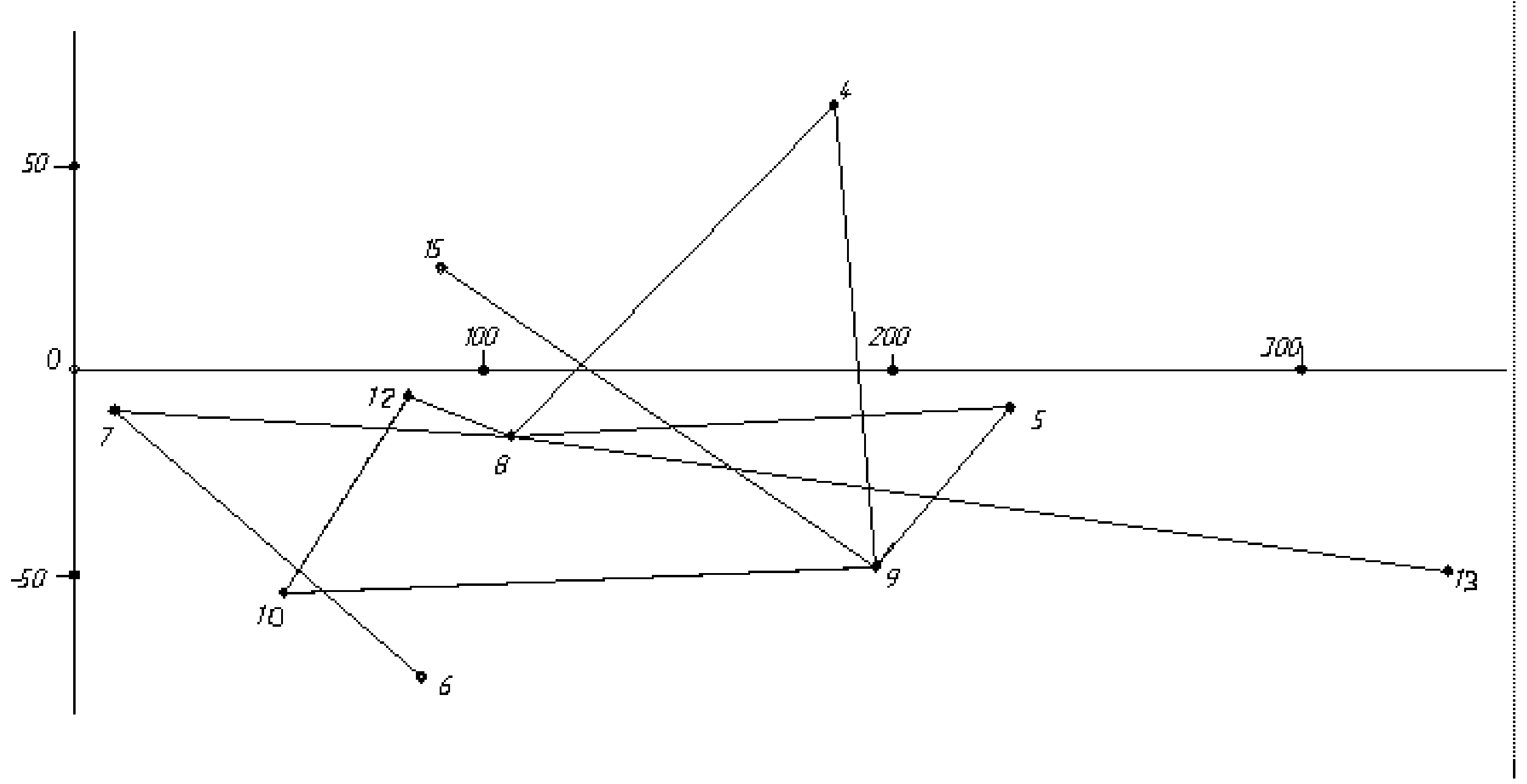

Fig. 8 Geometrical presentation of the list using (l,b) coordinates. (Kulikovsky, 1985)

So a natural question arises, when we study a different model of the Galaxy (new data for the dispersion of cosmic dust around the Sun), we see that the arrangement changes. How does the configuration change in this case?

| | | | |
|---|---|---|---|
| 1 | $\alpha$ Cen | | |
| 2 | Barnard's “runaway” star | | |
| 3 | Wolf 359 | 244 | 55 |
| 4 | +36 | 185.5 | 64.75 |
| 5 | $\alpha$ Cma | 229 | -9 |
| 6 | L726-8 | -176 | 75.5 |
| 7 | Ross 154 | 12.5 | -9.3 |
| 8 | Ross 248 | 110 | -17 |
| 9 | $\varepsilon$ Eri | 195.6 | -48 |
| 10 | Ross 128 | 270 | 59 |
| 11 | L 789-6 | 44.5 | -56 |
| 12 | $\varepsilon$ Ind | 336.3 | -48 |
| 13 | 61 Cyg | 82.5 | -5.6 |
| 14 | $\alpha$ Cmi | 219 | 13 |
| 15 | +59 | 90.5 | 24.5 |
| 16 | +43 | 116 | -20 |
| | | | |

The closest stellar systems ( according to their arrangement in 1999)

In this case the demonstrativeness of the figure remains. (fig. 36)

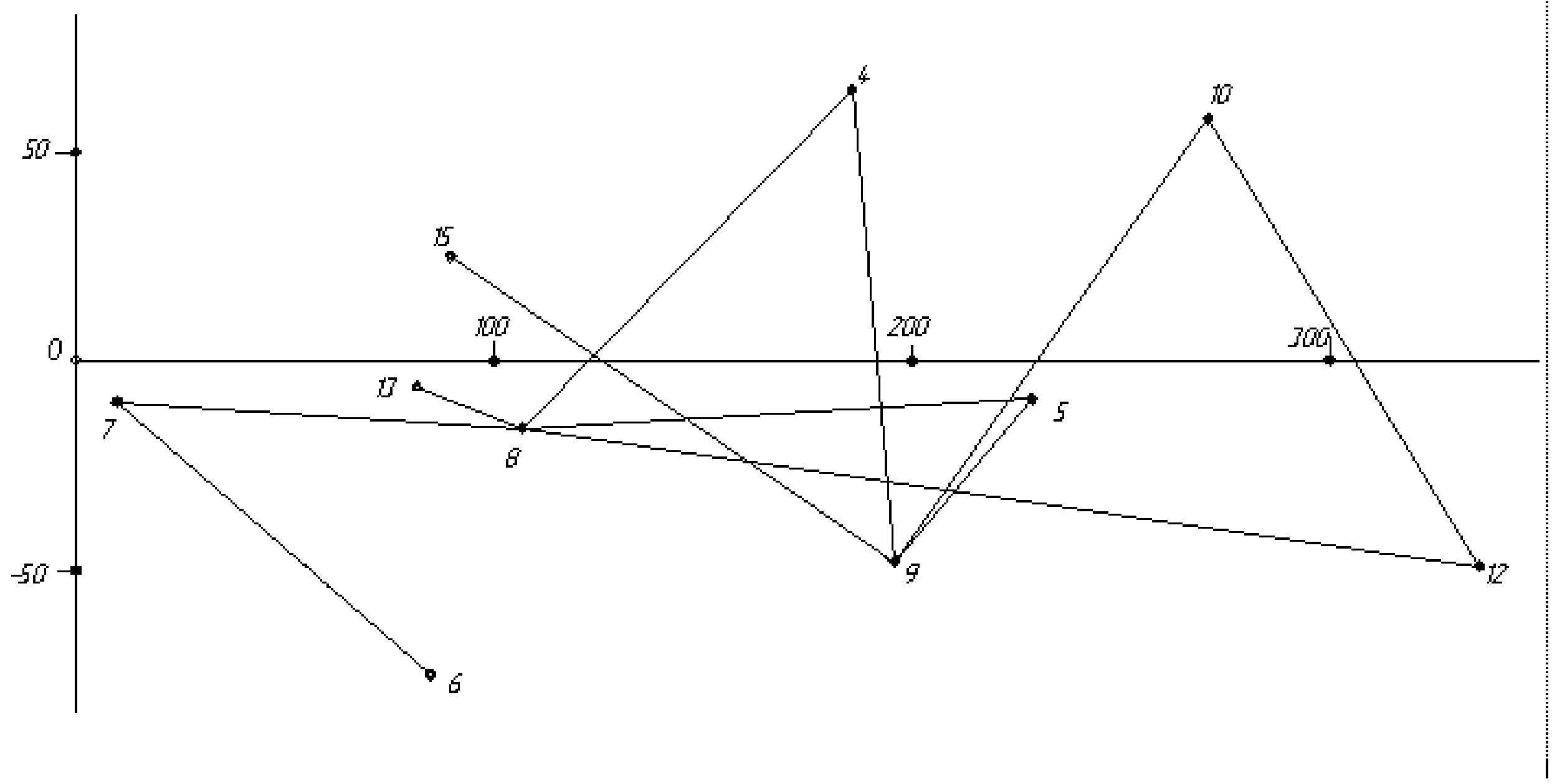

Fig.9 Geometrical presentation of the list using (l,b) coordinates.(1999)

Lets move on to the list of stars with the fastest apparent motion:

| | | | | | |
|---|---|---|---|---|---|
| 1 | Barnard's “runaway” star | $17^h\ 55^m$ ,4 | $+04^0\ 33'$ | | |
| 2 | Kapteyn's star | 05 09,7 | -46 00 | | |
| 3 | Groombridge 1830 | 11 50,1 | +38 05 | | |
| 4 | Lacaille 9352 | 23 02,6 | -36 08 | 6 | -65.2 |
| 5 | Cordoba 32416 | 00 02,5 | -37 36 | 347 | -75.5 |
| 6 | Ross 619 | 08 09,2 | +09 00 | 213.8 | 21 |
| 7 | 61 Cyg A+B | 21 04,6 | +38 30 | 185.5 | 64.75 |
| 8 | Lalande 21185 | 11 00,6 | +36 18 | 186 | 65 |
| 9 | Wolf359 | 10 54,1 | +07 19 | 244 | 56 |
| 10 | $\varepsilon$ Indi | 21 59,6 | -57 00 | 336.3 | -48 |
| 11 | Lalande21258 | 11 03,0 | +43 47 | 172 | 63 |
| 12 | 40 $o_2$ Eri | 04 13,0 | -07 44 | 200.5 | -38.2 |
| 13 | Wolf 489 | 13 34,2 | +03 57 | 330 | 64.1 |
| 14 | Proxima Centauri | 14 26,3 | -62 28 | 316 | 0 |
| 15 | $\mu$ Cas | 01 04,9 | +54 41 | 125.1 | -8.2 |
| 16 | Luyten's star | 07 24,7 | +05 26 | | |
| 17 | -15.4042 | 15 07,5 | -16 08 | | |
| 18 | $\alpha$ Cen | 14 36,2 | -60 38 | | |
| 19 | LP9-231 | 17 56,8 | +82 44 | | |
| 20 | Lacaille 8760 | 21 14,3 | -39 04 | | |

Stars with the fastest apparent motion near the Sun.

It turns out, that we get both the illustration of two notions: parallelism and incidenty (fig. 10):

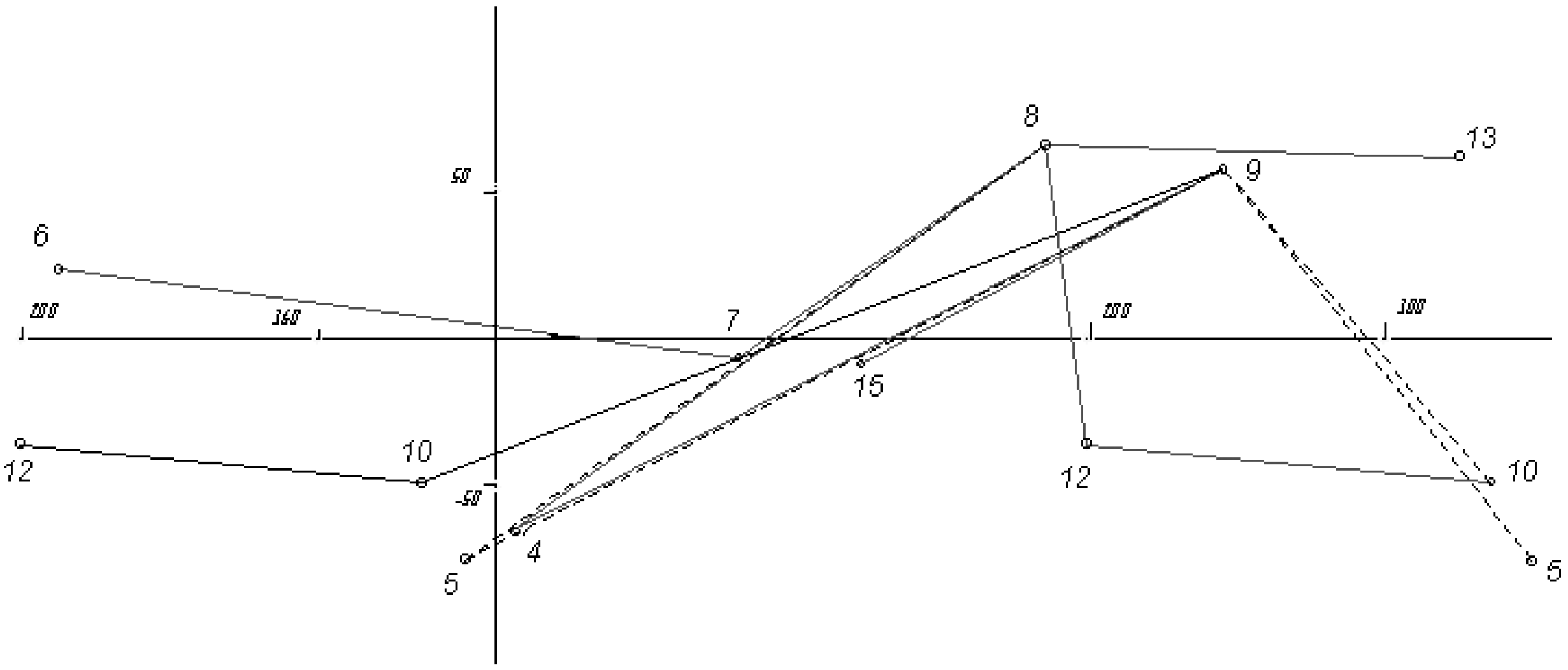

Fig. 10 Stars with the fastest apparent motion near the Sun.

Now, lets examine a list, that apparently has no relation to the location of the Sun and its surroundings. This is the list of the brightest clusters:

| | | | | |
|---|---|---|---|---|
| 1 | NGC 5139 | $3^m$.6 | 309 | 15 |
| 2 | NGC 104 | 4.1 | 306 | -45 |
| 3 | NGC 6656 | 5.0 | 10 | -8 |
| 4 | NGC 6205 | 5.7 | 59 | 41 |
| 5 | NGC 5904 | 5.8($M_V$ -$8^m$ .3) | 4 | 47 |
| 6 | NGC 6121 | 5.8 | 351 | 16 |
| 7 | NGC 6397 | 5.9 | 338 | -12 |
| 8 | NGC 2808 | 6.2(-9.5) | 282 | -11 |
| 9 | NGC 5272 | 6.2(-8.4) | 47 | 79 |
| 10 | NGC 6809 | 6.2(-7.2) | 9 | -23 |
| 11 | NGC 7089 | 6.3(-8.9) | 53 | -36 |
| 12 | NGC 7078 | 6.3(-8.7) | 65 | -27 |
| 13 | NGC 6341 | 6.4 | 68 | 35 |
| 14 | NGC 362 | 6.5(-8.2) | 302 | -46 |
| 15 | NGC 6254 | 6.5(-7.3) | 15 | 23 |
| 16 | NGC 6266 | 6.6 | 354 | 7 |

Clusters arranged according to V int ( Kulikovsky P. G. table Π 2.14).

Figure 11 correlates with the list above and with the first series of numbers:

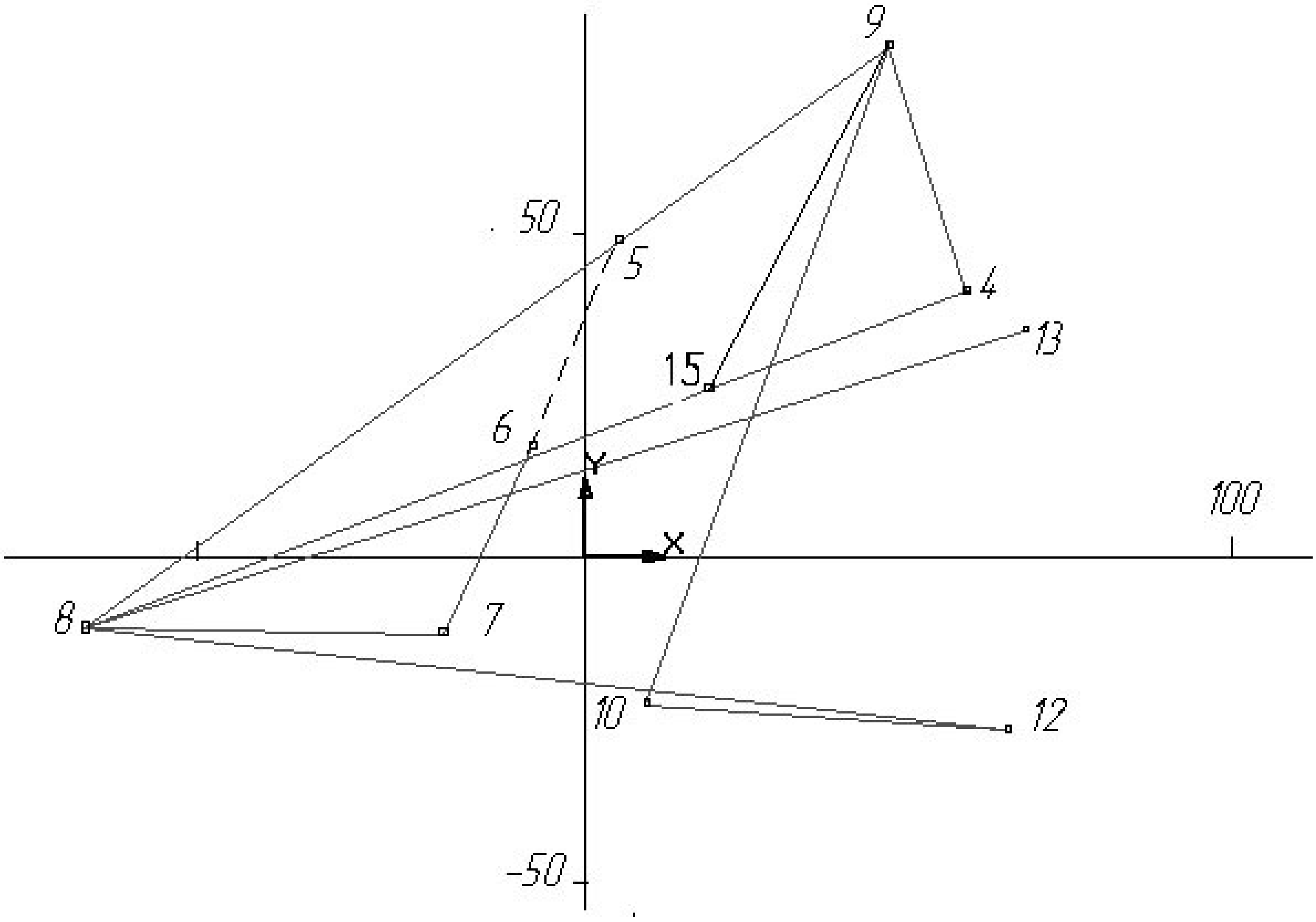

Fig.11 Bright clusters

Finally, lets study the ordered list of galactic open star clusters. Note that we can arrange them either according to radiosity or apparent diameter of the cluster. In both cases there is an illustration of geometric notions.

| | | | |
|---|---|---|---|
| 4 | Coma | 216 | 82 |
| 5 | NGC 6475 | 356 | -4 |
| 6 | NGC 3532 | 289 | 2 |
| 7 | NGC 2632 | 194 | 35 |
| 8 | NGC 2682 | 216 | 31 |
| 9,10 | NGC 869,NGC 884 | 135 | -4 |
| 11 | NGC 3114 | 283 | -3.7 |
| 12 | NGC 6405 | 356 | -0.2 |
| 13 | NGC 3766 | 294 | 0 |
| 14 | NGC 2287 | 230 | -10 |
| 15 | NGC 4755 | 303 | 3 |

Open star clusters arranged according to radiosity.

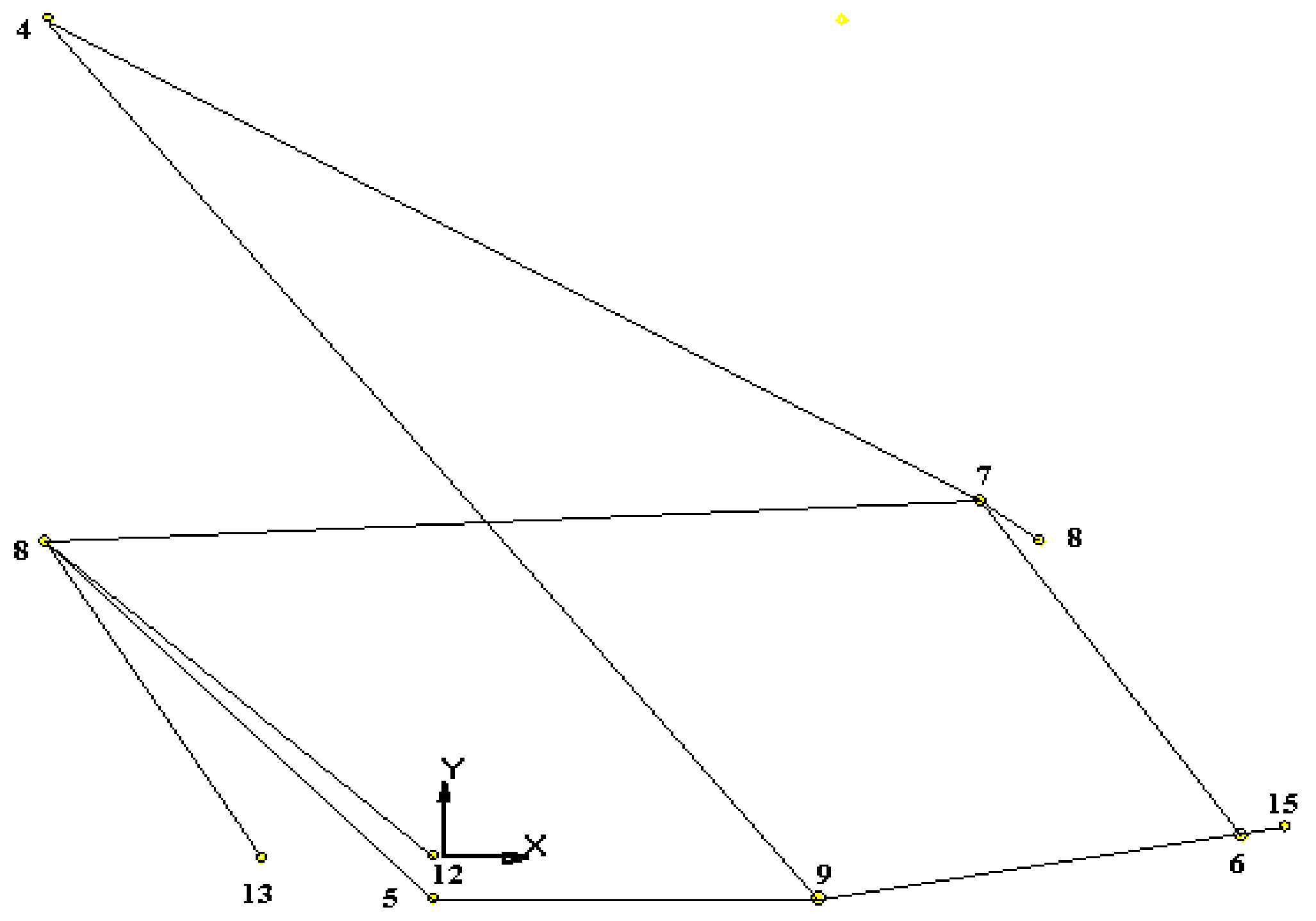

Fig. 12 Open star clusters arranged according to radiosity.

| | | | |
|---|---|---|---|
| 4 | Coma | 216 | 82.1 |
| 5 | Pleiades | 166 | -3.6 |
| 6 | NGC 3532 | 289 | 2 |
| 7 | NGC 6475 | 356 | -3.8 |
| 8 | NGC3114 | 283 | -3.7 |
| 9 | NGC 6405 | 356 | -0.2 |
| 10 | NGC 2287 | 230 | -10.2 |
| 11 | NGC 1039 | 143.5 | -16 |
| 12 | NGC 2168 | 186.5 | 23 |
| 13,14 | NGC 864,869 | 134.7 | -3.8 |
| 15 | NGC 6494 | 9.9 | 2.7 |

Open star clusters arranged according to apparent diameter.

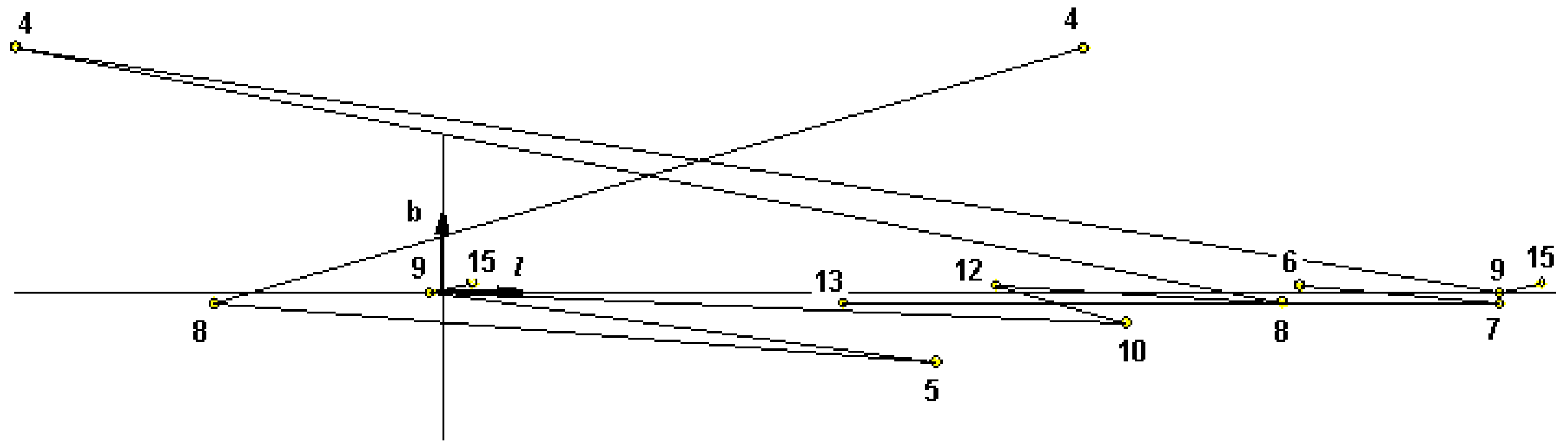

Fig. 13 Open star clusters arranged according to apparent diameter.

We may now move on from the observations to mathematical problem stating:

From the set of N numbers, corresponding to K planes with sets of points, select the largest subset of numbers and the order of their arrangement, in which simultaneously on all K planes such concepts as parallelism and incidenty can be geometrically illustrated.

In pure formulation this is a difficult problem of multi-criteria optimization, because so ordered lists should be examined. How can we correlate measures of the closeness of points to the lines and measures of parallelism of these lines?

If it turns out that the set of numbers 15,9,4,8,13,8,12,10,9,5,8,7,6 is the best suitable for all the lists we considered above, then the probability of getting the same set of numbers approximately equals to 1/19!, which is highly doubtable, but a skeptic can still deal with it. It's important to mention that the search of such sets (15,9…,etc.) can be conducted without straightforward enumeration, if the problem of clusterization is previously solved. The angle included between two intervals can be chosen as metrics.

Then, lets consider that the galactic coordinate system is a field or a "board" where the information will be presented. Lets also examine the 3$^{rd}$ series of numbers registered by Stormer, and Van der Pol with delay time of 12, 5, 8.

After analyzing different arrangement of stellar magnitudes, which were measured using photoelectric method (the data was taken from Rulikovsky's monograph), we get the following stars that correspond to the set of numbers: 5 – Alpha Centauri (the Arabs call it Toliman), 8 – Procyon, 12 – Altair. It turns out that these three stars are on a line, and point 5 is an intersection point of 2 lines (if we consider cylinder).

The study of the second series of Stormer 12, 14, 14, 12, 8 shows, that there is another star which adds to the three ones mentioned above. This star is Aldebaran.

After examining other lists of stars, we may conclude that for some of the lists, number triples 12, 14, 8 and 12, 5, 8 are approximately on one line and they form an intersection. These are the lists of stars with the fastest apparent motion (fig. 14), the closest stars to the sun (1985 and 1999), (fig. 15), (fig. 16), the most luminous stars and the list of the brightest clusters(fig 17).

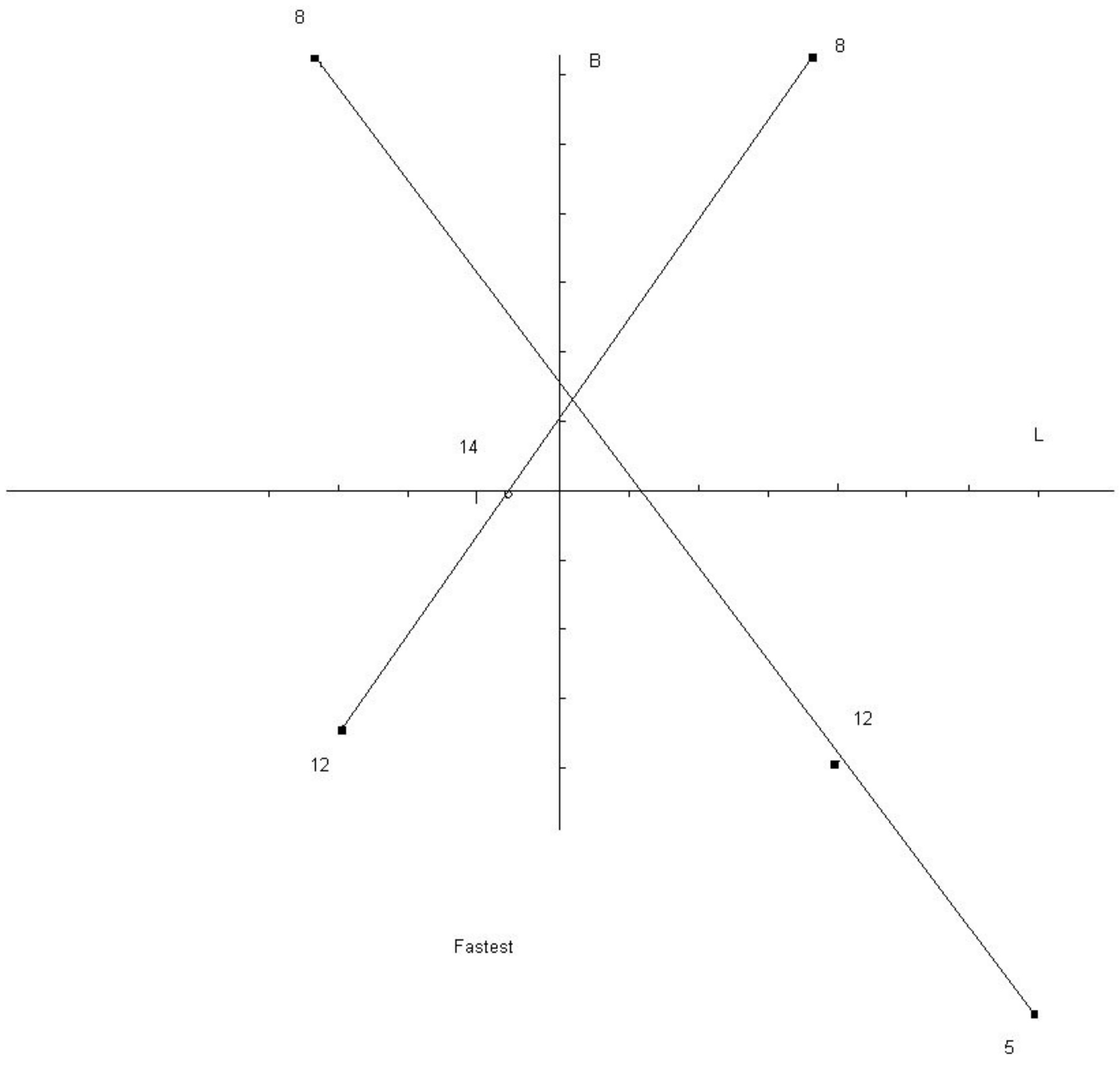

Fig.14

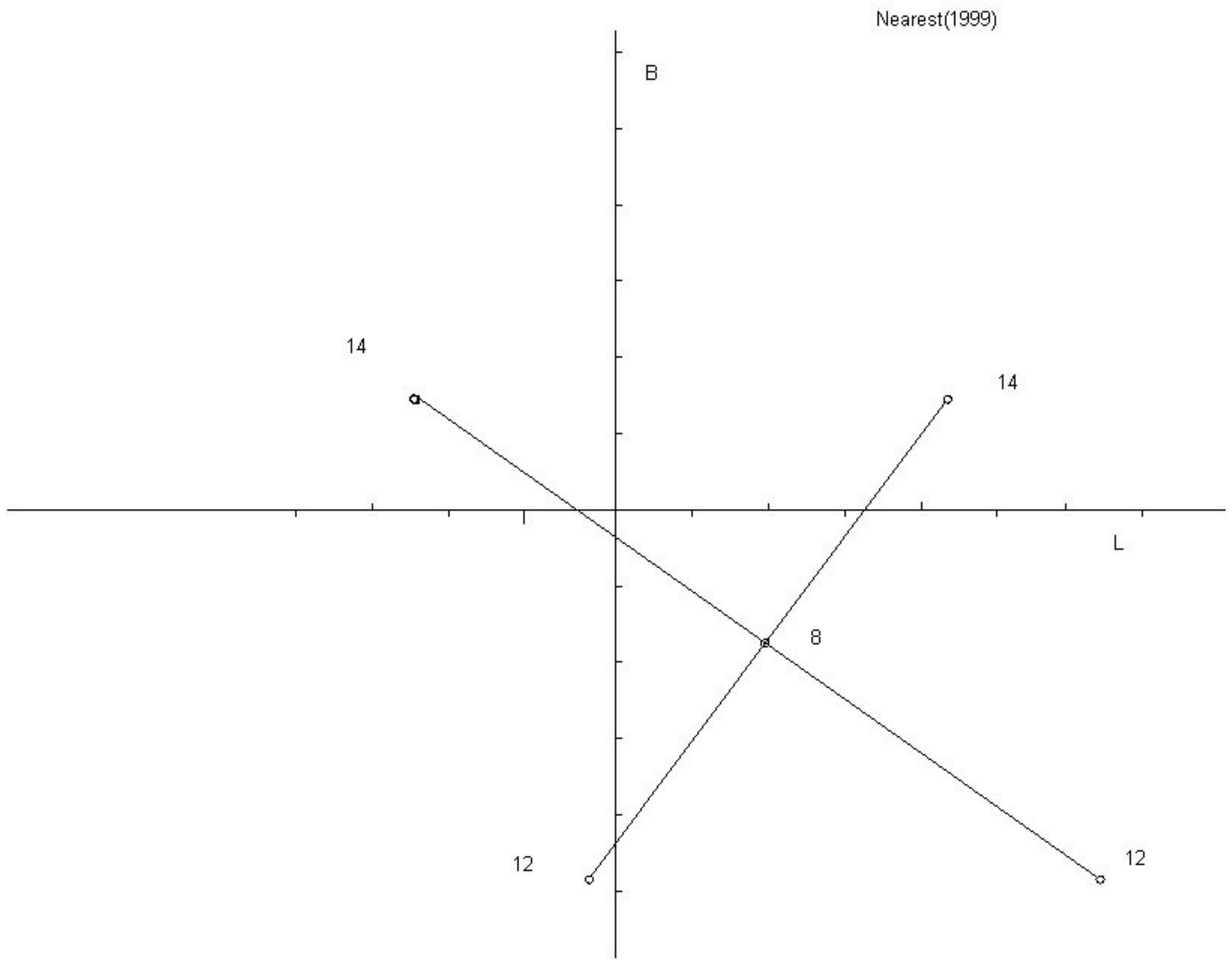

Fig. 15.

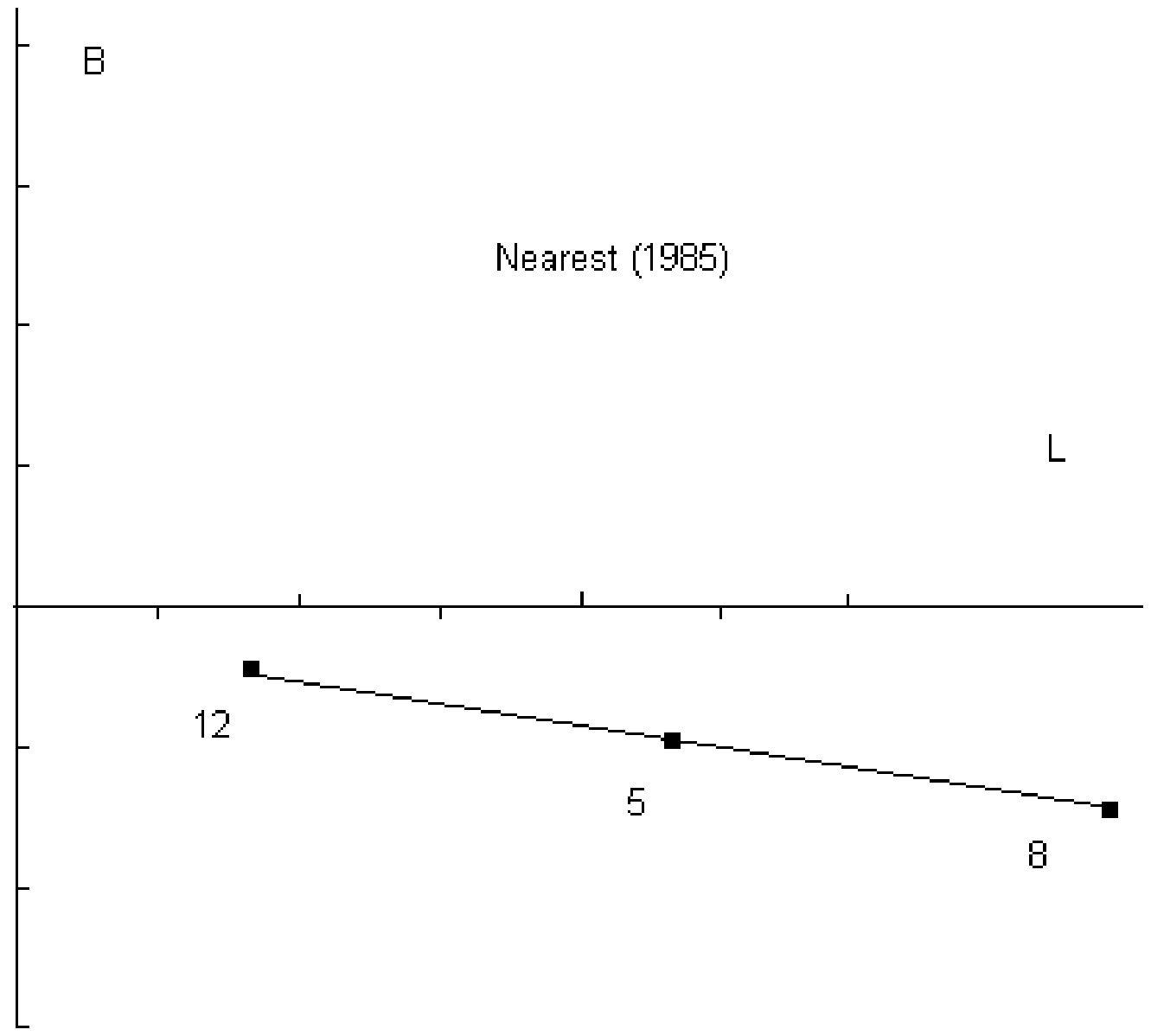

Fig. 16.

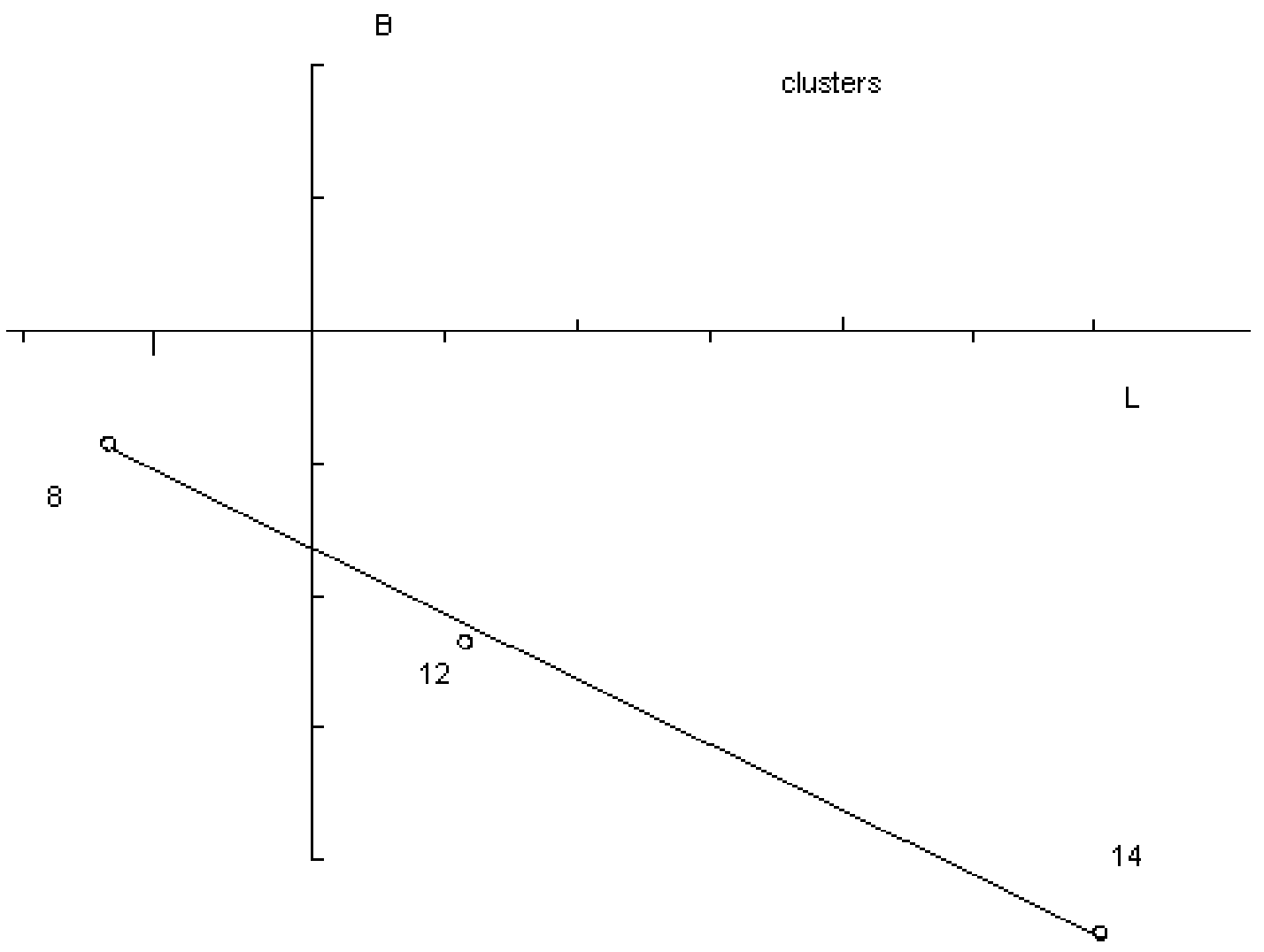

Fig. 17.

Simple search of numbers demonstrates that this two sets of numbers 12, 8, 14 and 12, 8, 5 are unique, because they lie both on one line and on the intersection. The possibility of independent and random choice of these very sets approximately equals $\sim (1C_{19}^{3})^{2}$ , which is about $10^{-6}$ .

Suppose, that not only delay time of echoes is important, but also the complements to 20. We consider the conditions of the experiment performed by Stormer and keeping in mind that these signals are artificial, we can study both delay time and their complements as equivalent data. (fig. 18)

Two new stars are being added 6 – Capella and 15 – Antares. Now the figure becomes complete and absolutely symmetrical. Can this be a simple coincidence? Even though the transition to another spherical coordinate system, ecliptic (Fig. 19) and equatorial (Fig. 20), preserves the symmetry. On the other hand, is it possible that this dissimilarity from the ideal straight lines makes these properties very possible?

Pay attention to the fact that the number 20, or, in other words interval between the signals sent form Earth is “presented” on these pictures.

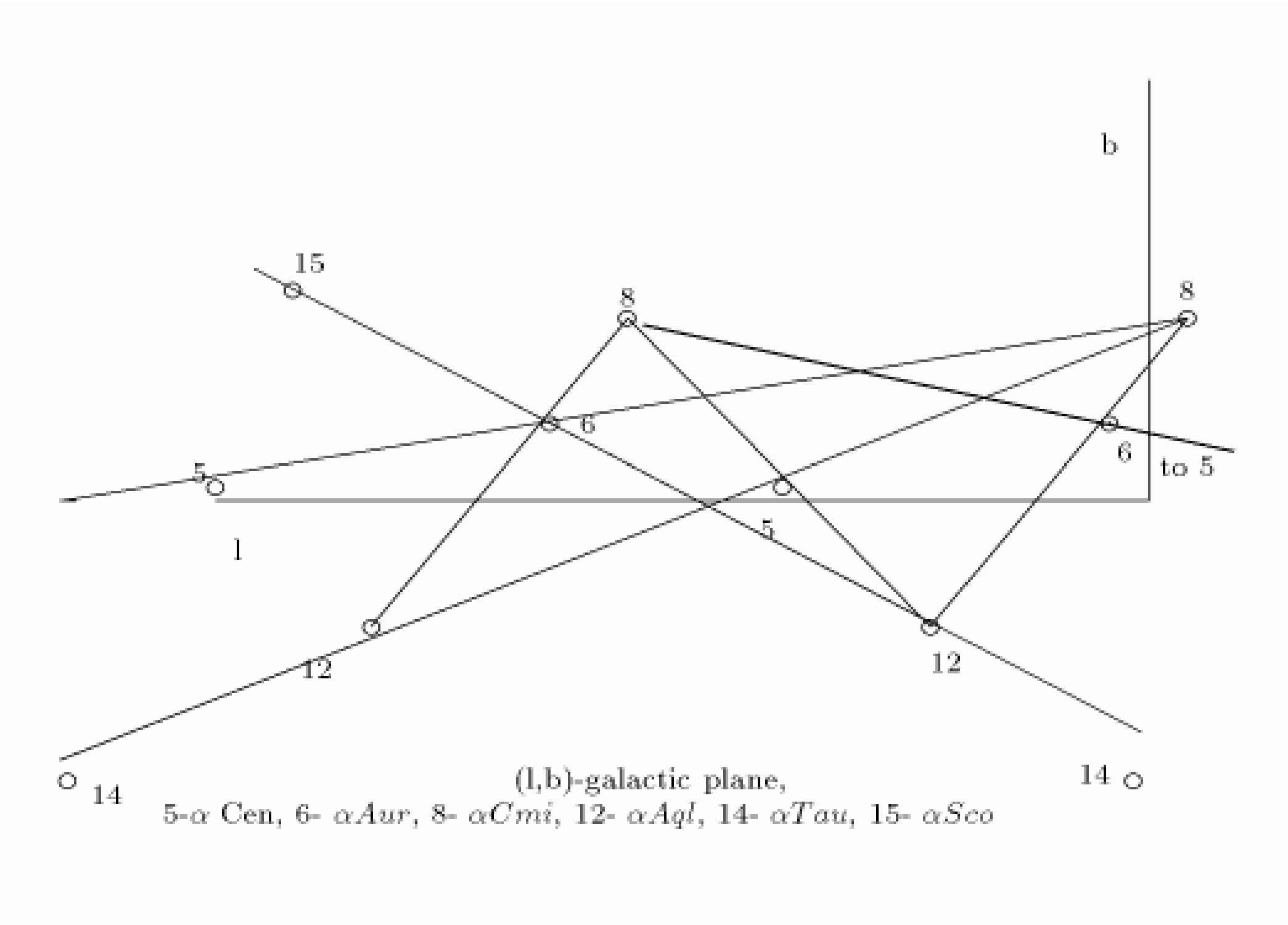

Fig. 18 Galactic plane

For example, in galactic model representation the points with numbers 5 and 6 lie on the concurrent lines, which connect points with numbers in sum giving 20. In the equatorial coordinate system concurrent lines contain points, the numbers of which are mutually complementary to 20. If we consider all the complements to 20, for the first series of Stormer, then we again get the illustration of parallelism. If we connect the points with numbers 5 11 16 12 7 12 8 10 11 15 12 13 14, that are the complements of numbers 15 9 4 8 13 8 12 10 9 5 8 7 6, then the "reversed image" we get strikes us with its symmetry! (Fig. 21) Again we get the illustration of the parallelism.

Numerical experiments show, that it is quite difficult to get such symmetries, but studies of independent data is even more evidential. The data was taken from the results of the experiments of Appleton in 1934.

It turns out that for 50 points (positions of 50 stars) we can already distinguish invariant configuration " a network": the equatorial model representation(fig. 22), ecliptic model (fig. 23), galactic model(fig. 24). In all the models each delay time corresponds to a star with specific number and the property, which we named "lie on the same line" is confirmed. Note, that the network is formed with the help of two elements of cylinders.

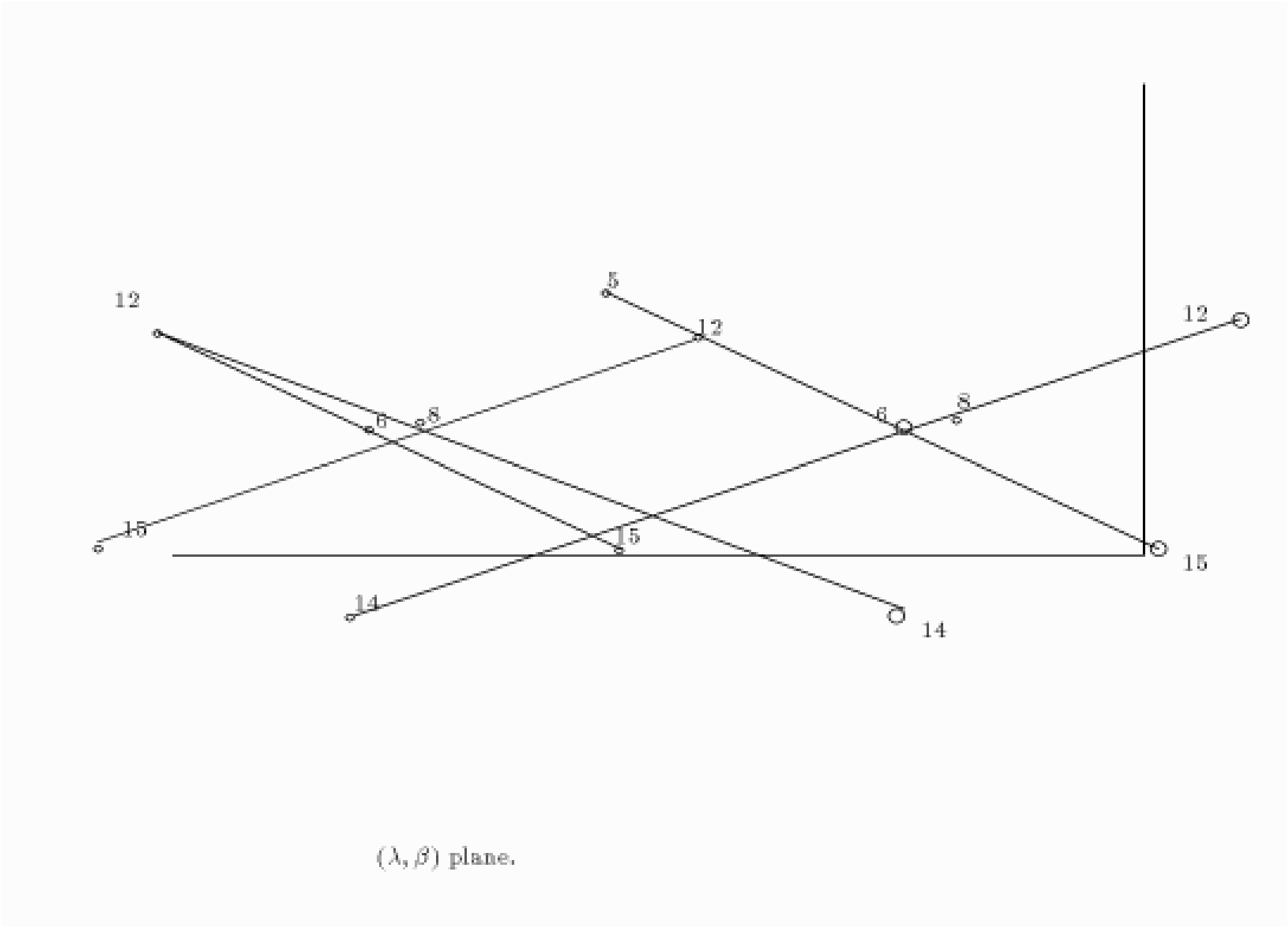

Fig. 19

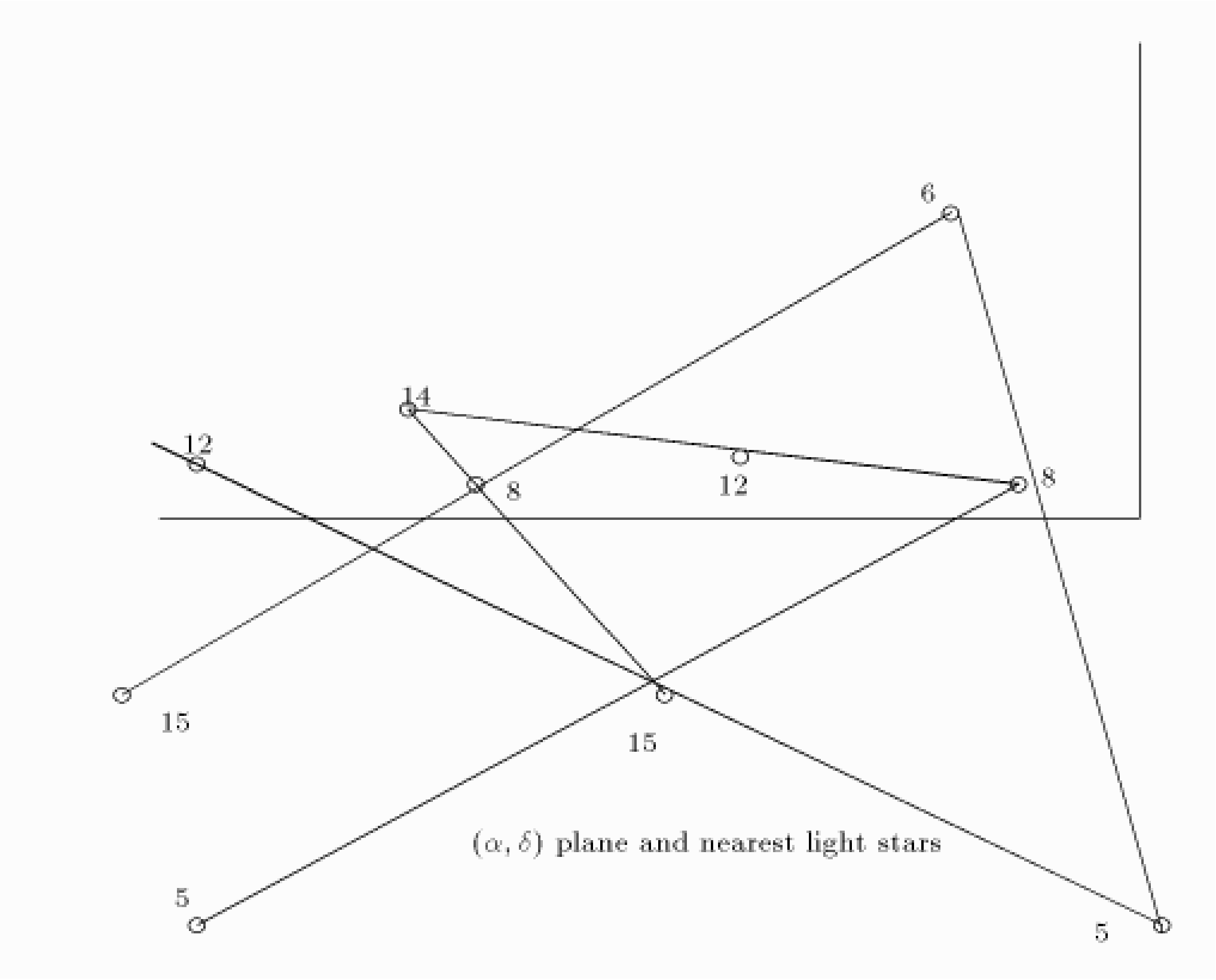

Fig. 20.

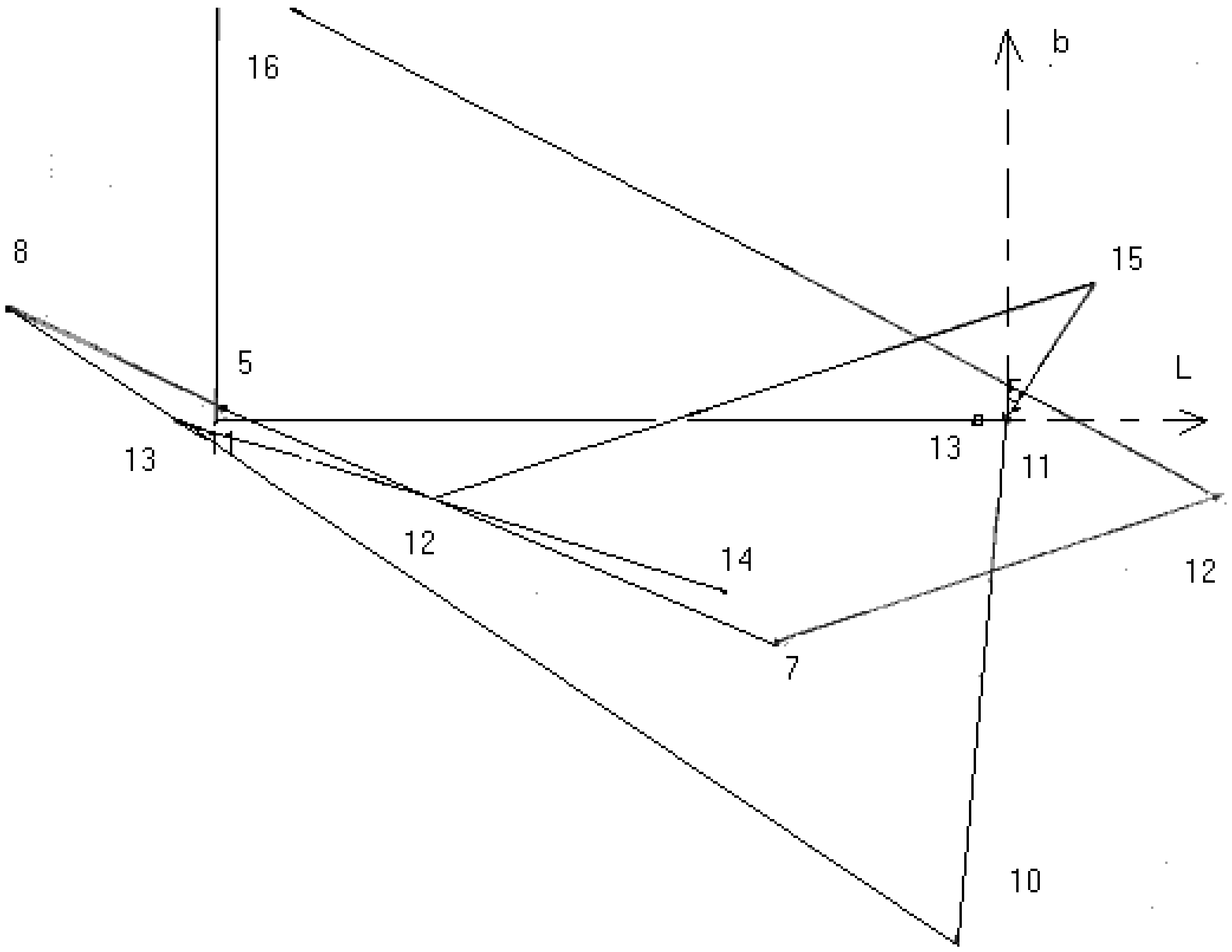

Fig. 21.

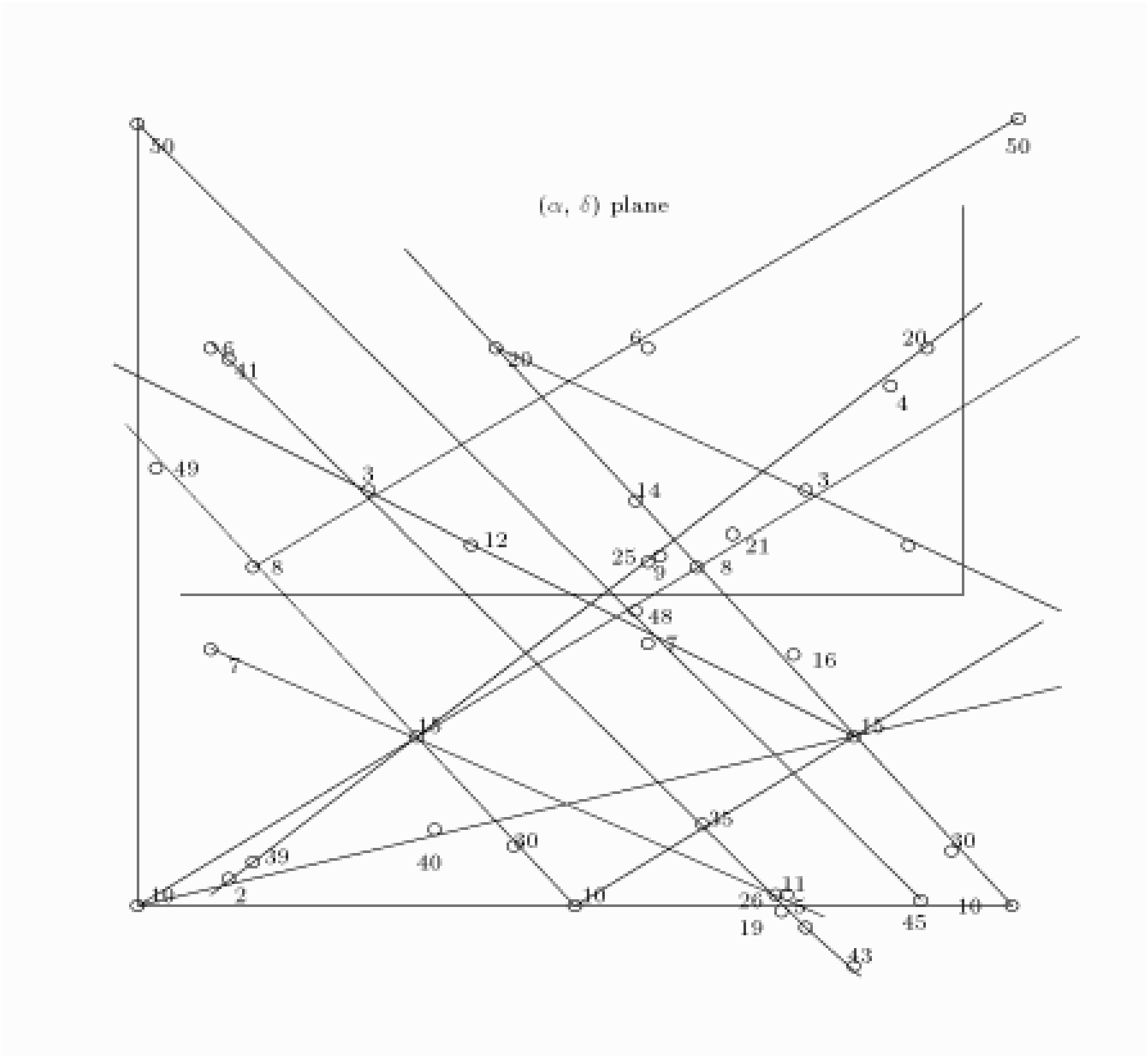

Fig. 22.

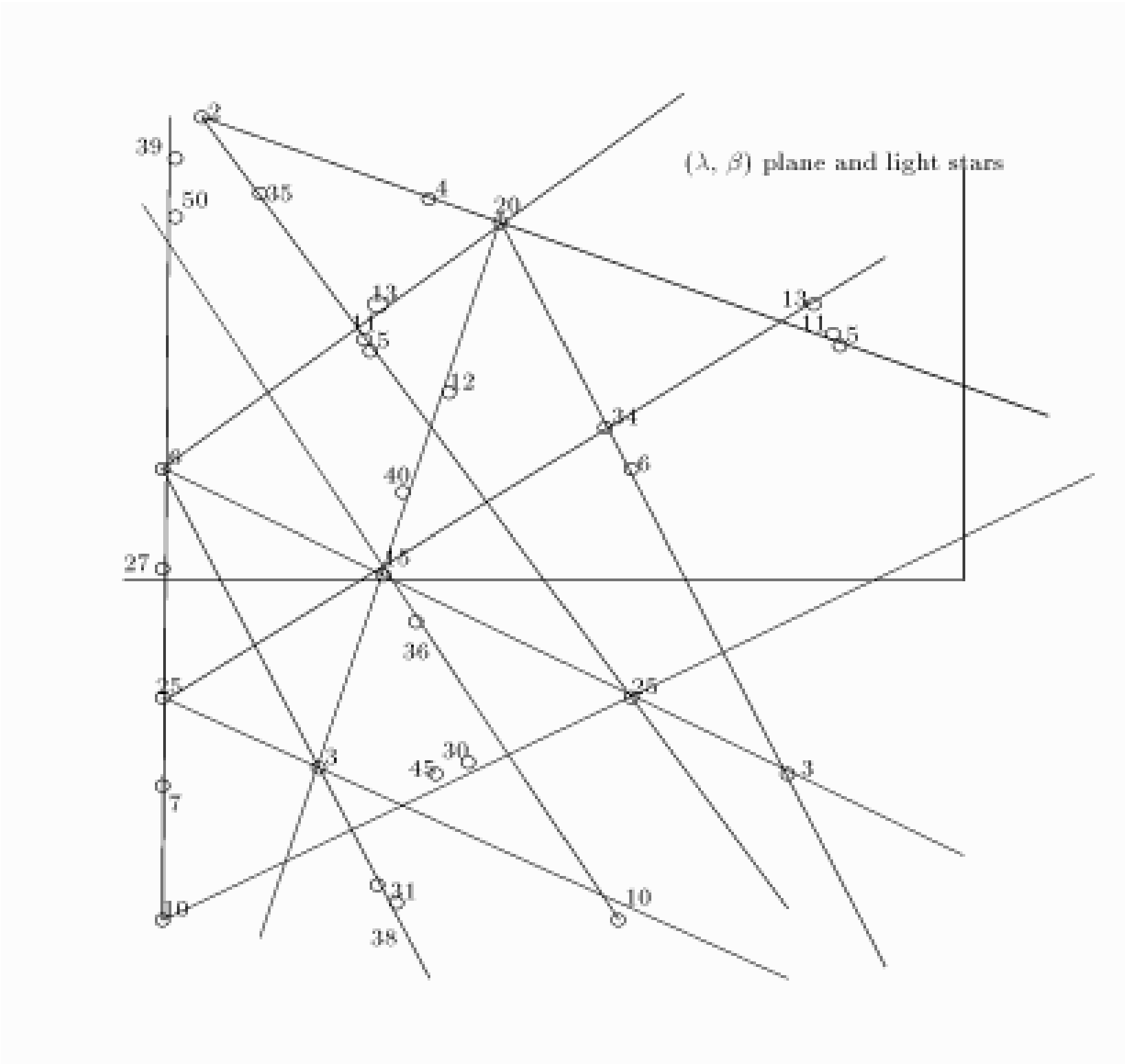

Fig. 23.

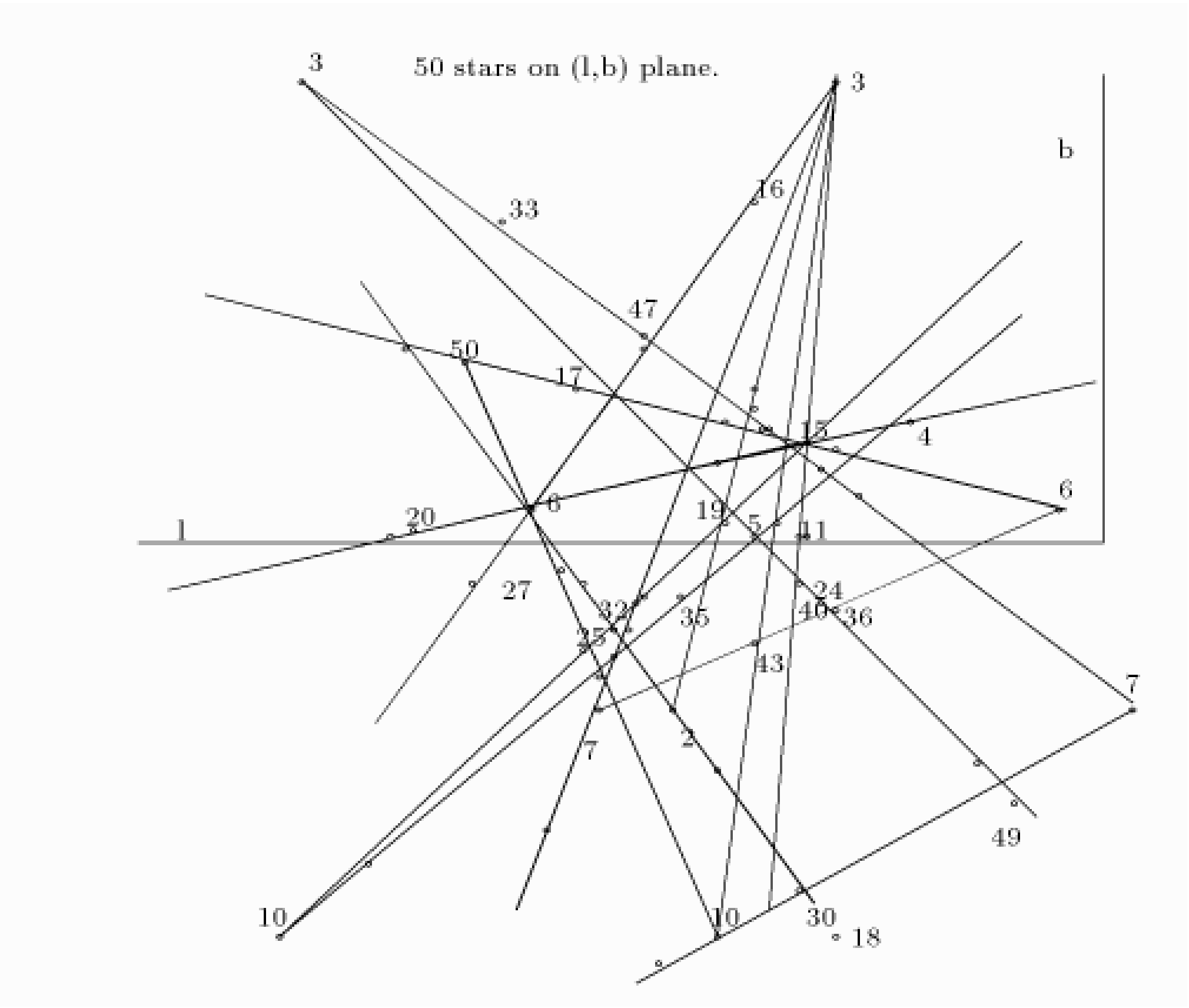

Fig. 24.

It turns out that the 4th series of Stormer is also the illustration of the property of parallelism. For example, if we trace the point on the plane (*l,b*) in the following order 12 8 5 14 14 15 12 7 5 5 13 8 8 8 13 9 10 7 14 6 9 5 9, we get the same figure. (Fig 25).

From now on,  we will consider, this large series of numbers as "cover" of a letter, which also fulfills a function of "content". We'll also try to find some objective explanation to how can randomly distributed points from so many almost parallel lines? The results of closer examination show that if 50 coordinates of the closest stellar systems or systems with the fastest apparent motion are taken, then there's almost no parallelism demonstrated. The same thing happens when we try to examine sets of randomly chosen points.

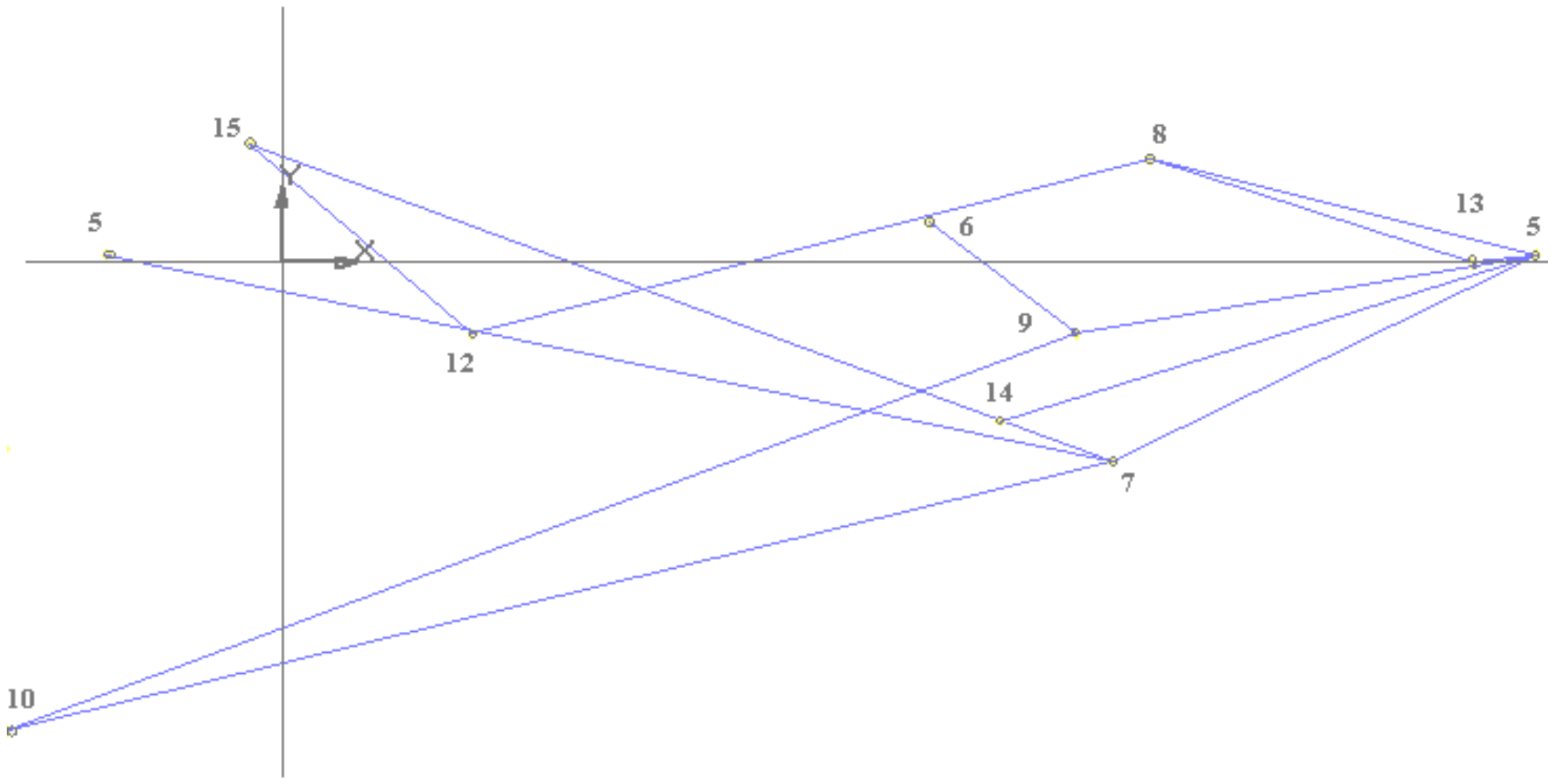

Fig. 25 Representation for 4th series

So what could this mean? What kind of hints are behind these figures?

First of all, can we talk about the figures in a more accurate language – the language of mathematics? Is there anything that may look alike in the language of geometry?

There are some mathematical objects that may "look like" natural geometrical ratios. This objects are called configurations[9]. The most basic example of this configurations is Brianchon–Pascal configuration.(Fig. 26).

The data of 9 points satisfy the following conditions: three lines pass through each point, there are three points on each line. We formally call this object (9)_3.

Note, that similarities in the properties may occur when similar operations are performed. Configurations are mathematical objects that are basic for projective geometry.

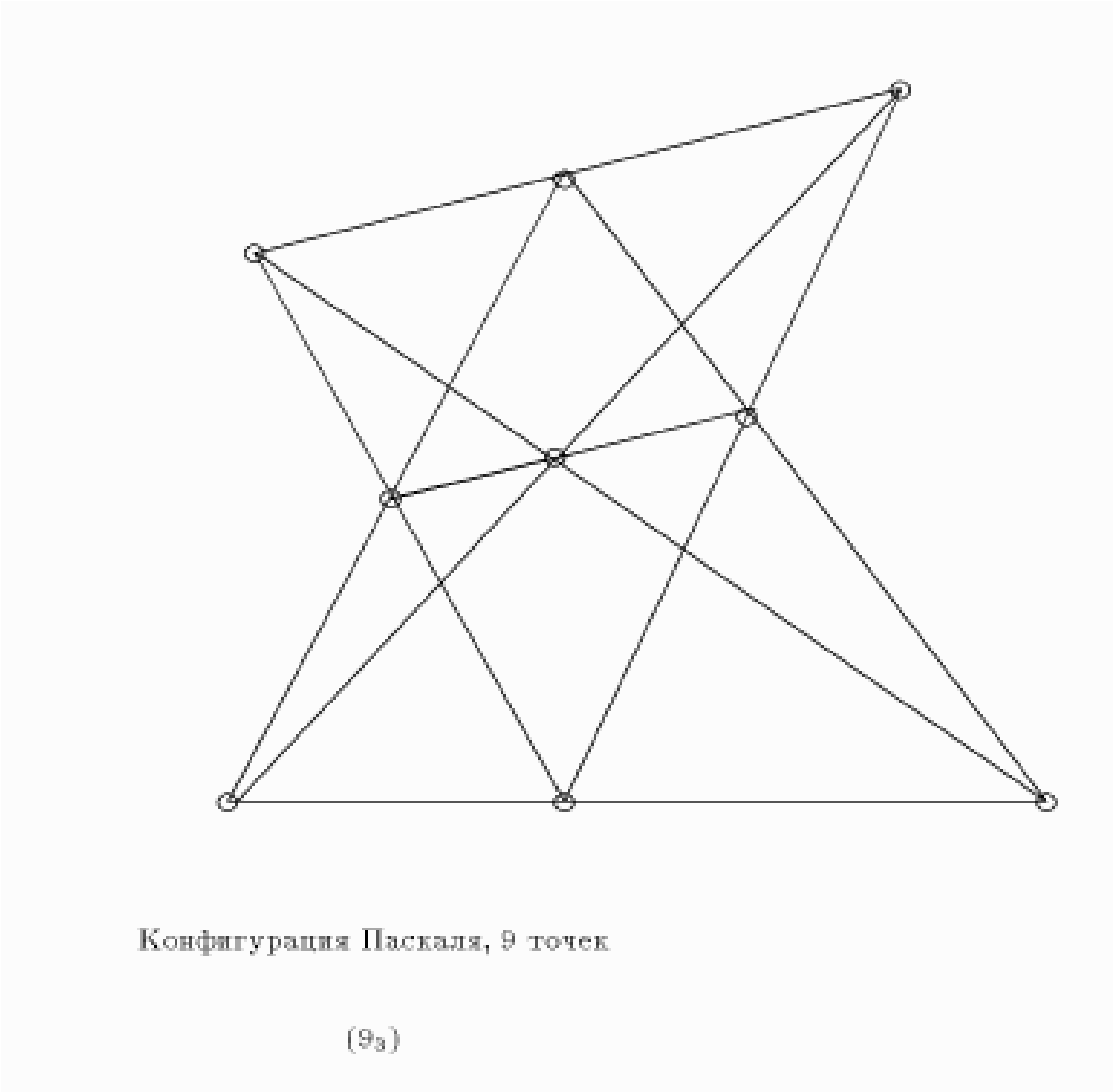

Fig. 26 Pascal configuration

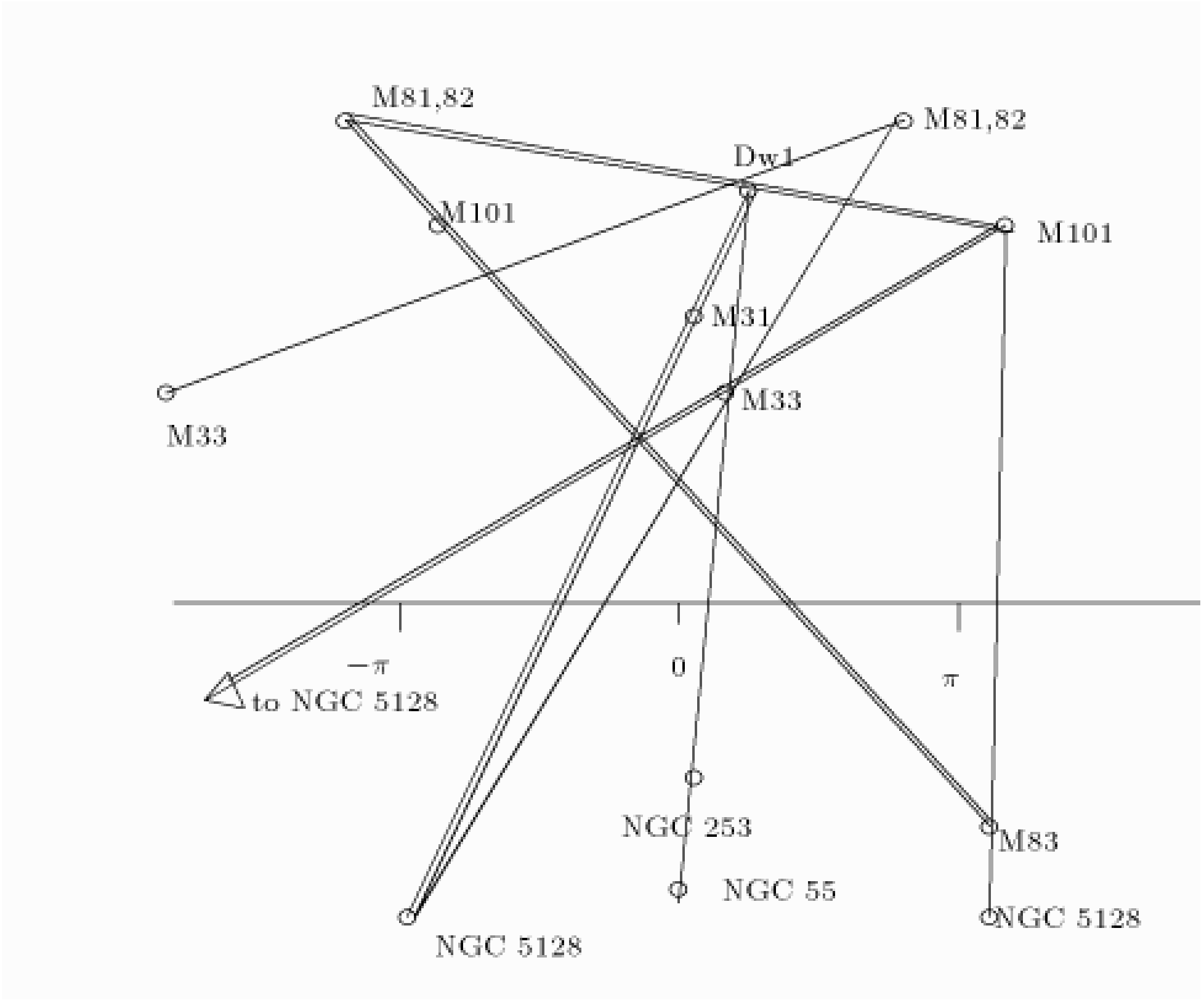

Fig. 27

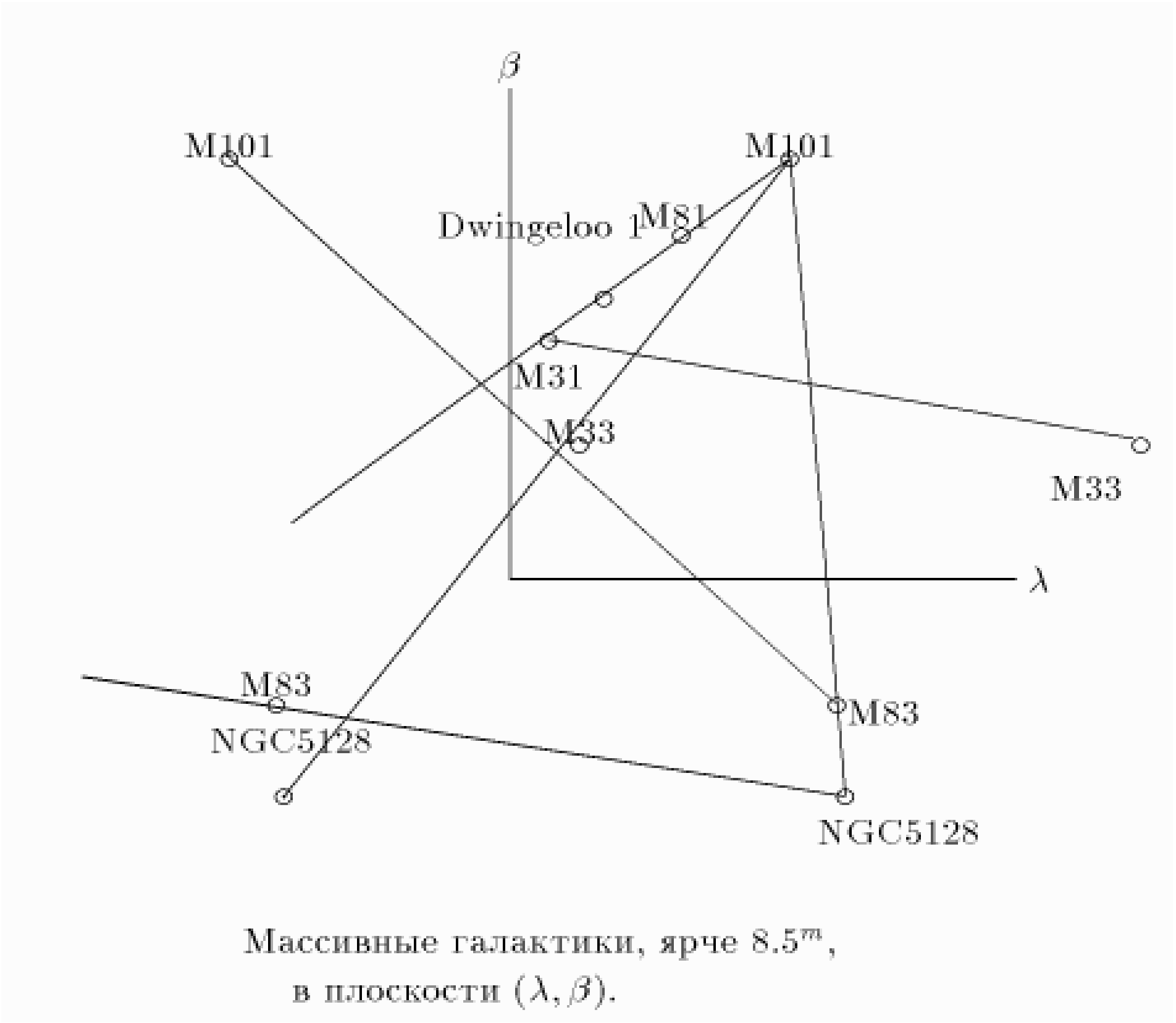

Fig. 28

Lets study the representations of the nearest  supergiant galaxies.

We examine galaxies that can be compared according to their masses. For example, Great Andromeda Nebula, our Galaxy, Dwingeloo 1, Triangulum Galaxy. They are giant galaxies. According to their mass dwarf galaxies are thousand and ten thousand times smaller then giant galaxies.

Based on the construction of graphs we used for the equatorial coordinate system, we get the analog of Pascal configuration. (Fig. 27) But no arrangement according to their luminosity is presented.

As in the case of the closest stars, there is also some arrangement in ecliptic (Fig. 28), galactic (Fig. 29) and supergalactic(Fig. 30) coordinate systems. All the coordinates of galaxies were taken from Extragalactic Database (NED):http://nedwww.ipac.caltech.edu/index.html

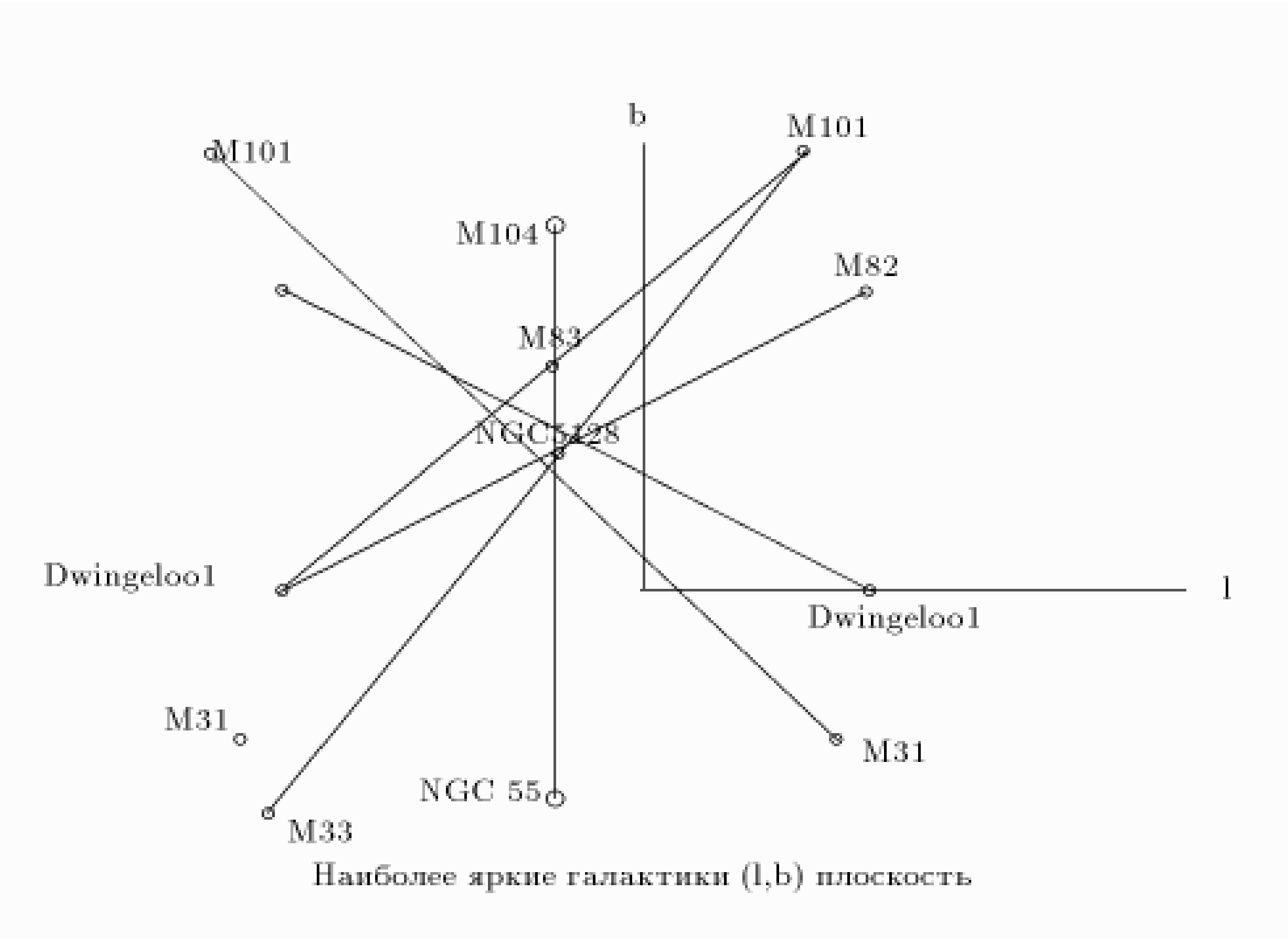

Fig. 29 The brightest galaxies (l, b) plane

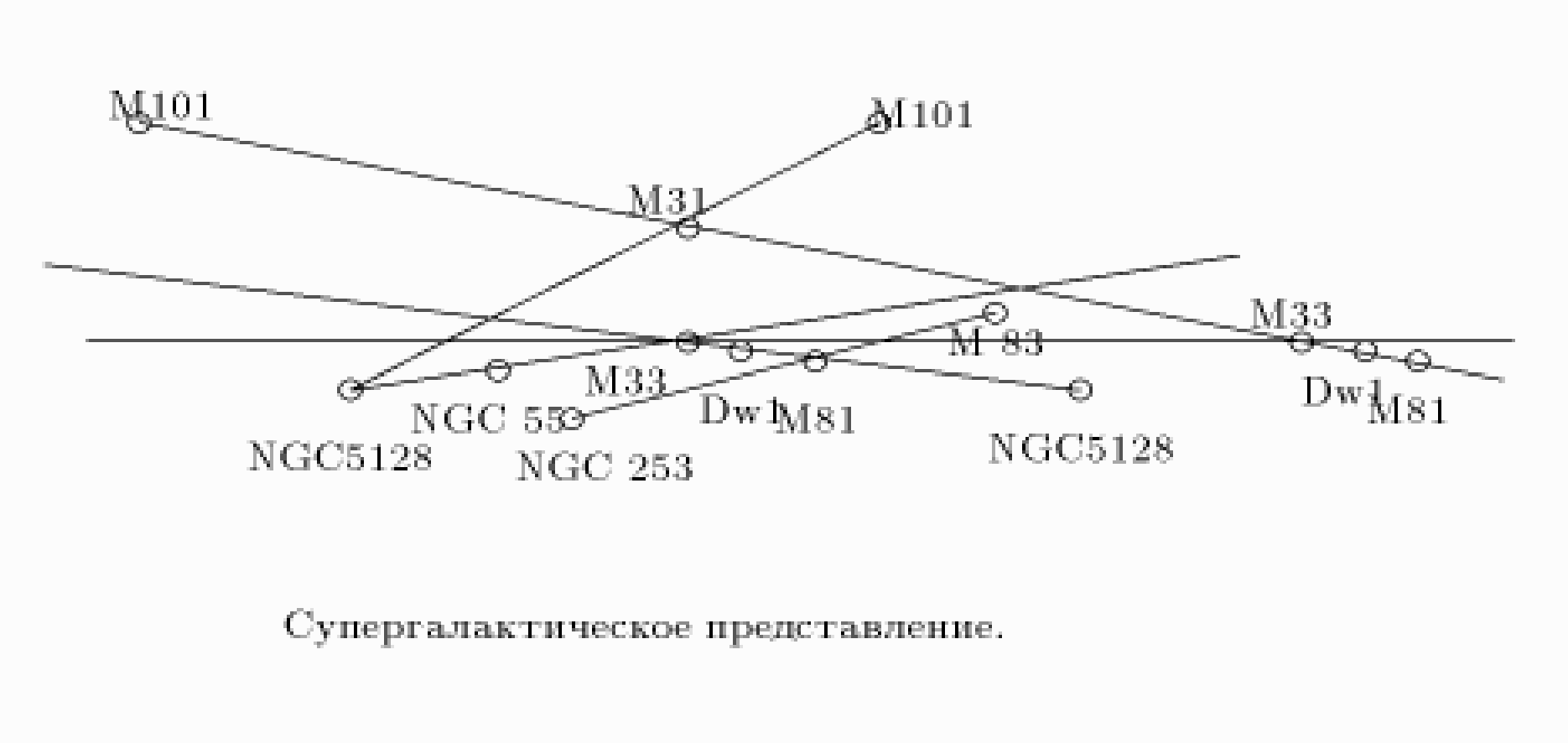

Fig. 30 Supergalactic representation.

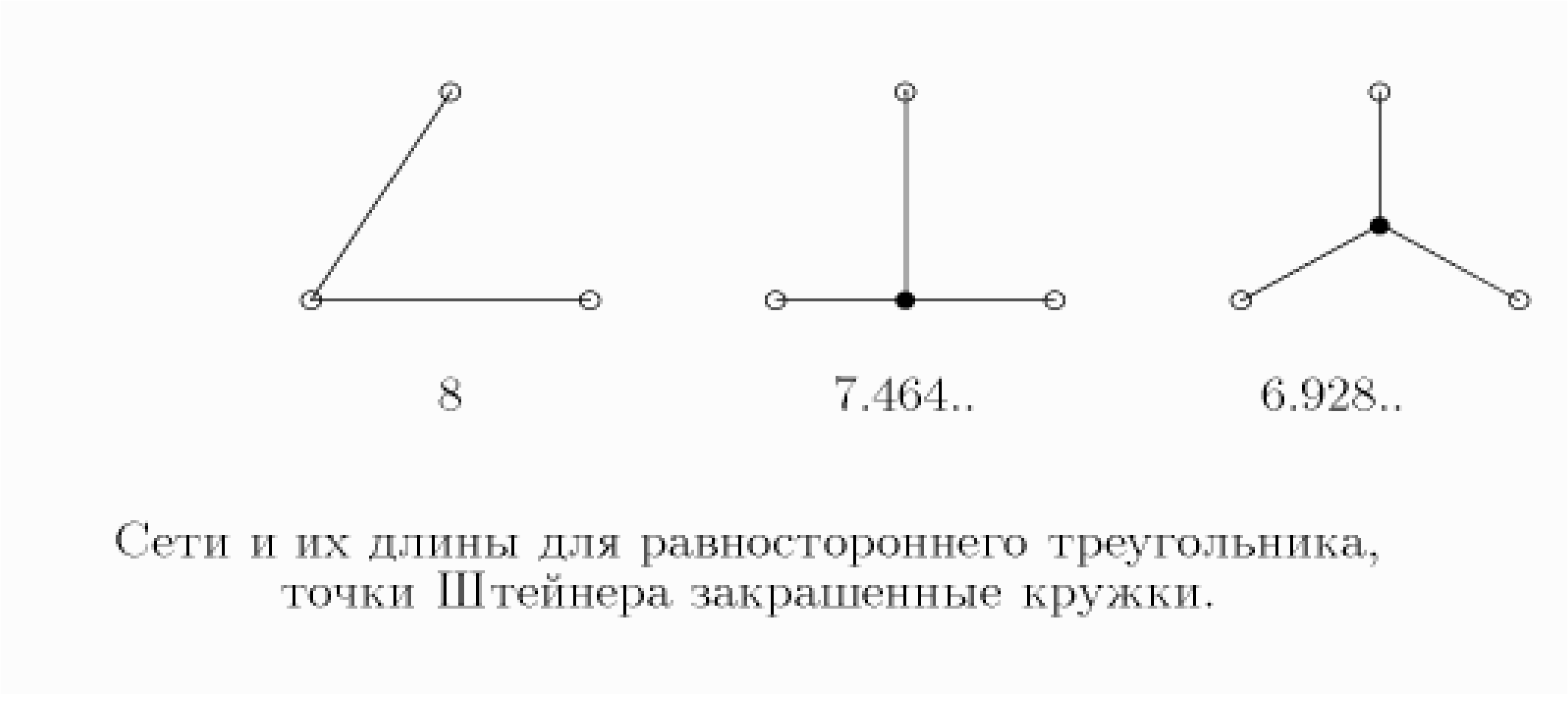

Fig. 31 Networks and their lengths for equilateral triangle, filled points are the

points of Steiner

The properties, which we determined first for the position of stars and then for the galaxies in a way clarify the situation. Galaxies remain almost still in relation to each other. Their velocities are negligibly small compared to the distances among them. It can be stated that total energy almost coincides with potential energy. But this means that potential energy reaches its maximum value.

Can it be that configurational properties are the result of the experimental state of the system?

What can we say then about systems of the closest most luminous and massive stars? Objects that are close to the Sun are located in what is called corotation circle [14], where the rotational velocity of the Galaxy coincides with the velocity of spiral arms. This means that for millions of years the surroundings of the Sun move outside the zones where nonstationary process takes place. During this period masses were distributed and the most luminous and massive stars were stably formed.

All the above mentioned leads us to the examination of a well-known mathematical problem, which is called Steiner's tree problem:

N points are given in the plane or in the space. The goal is to connect them by lines of minimum total length. Note, that besides given points there also can be some additional ones.

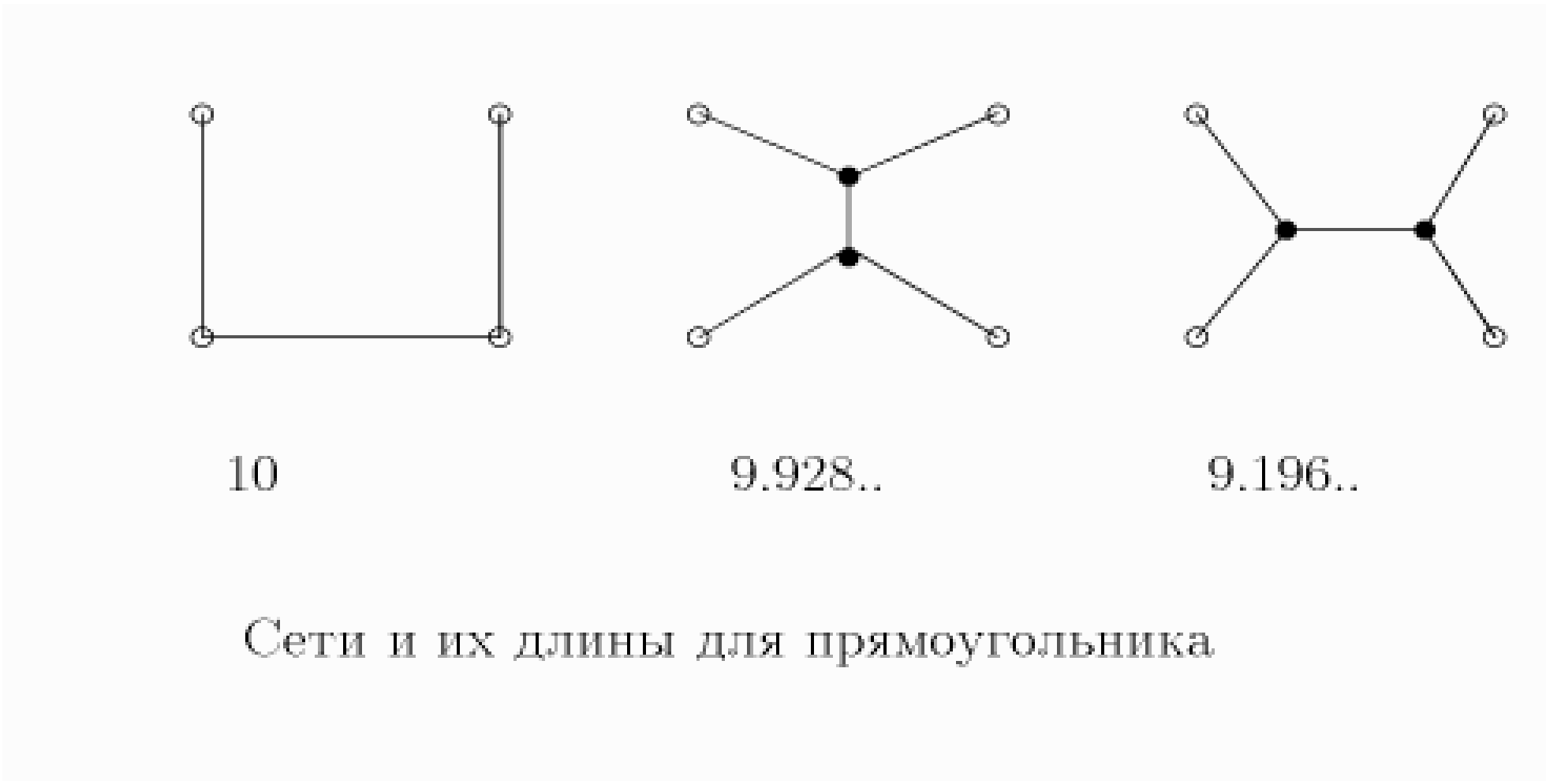

Fig. 32 Networks and their lengths for a rectangle

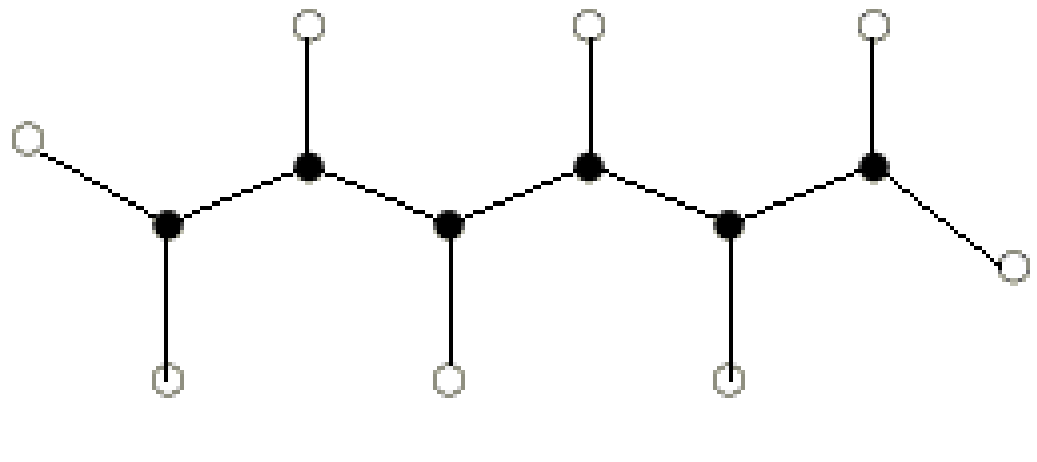

Fig. 33 Exact solution

The problem, in it's general form, first appeared in the article written by Milos Kessler and Wojciech Jarnik published in 1934, but the problem itself does not become widely known until 1941, when Richard Courant and Herbert E. Robbins put it in their book "What is mathematics? ". Courant and Robbins linked this problem to the work of Jakob Steiner, a 19-th century German mathematician. But the traces of the problem go back to the times of Evangelista Torricelli and Bonaventura Cavalieri, who solved a special case of this problem: find the point P, the sum of the distances from which to each of the three given points is minimal. They deduced that if the angles at point P are all 120 degrees then the total distance is minimized. They also proposed the algorithm for the solution of the problem.

To illustrate all the above mentioned, lets examine geometrical figures, that can be treated as solutions for the Steiner's tree problem for small number of points. For the case when there are only three points that are the vertices of a triangle the description and solution are shown on figure 31.

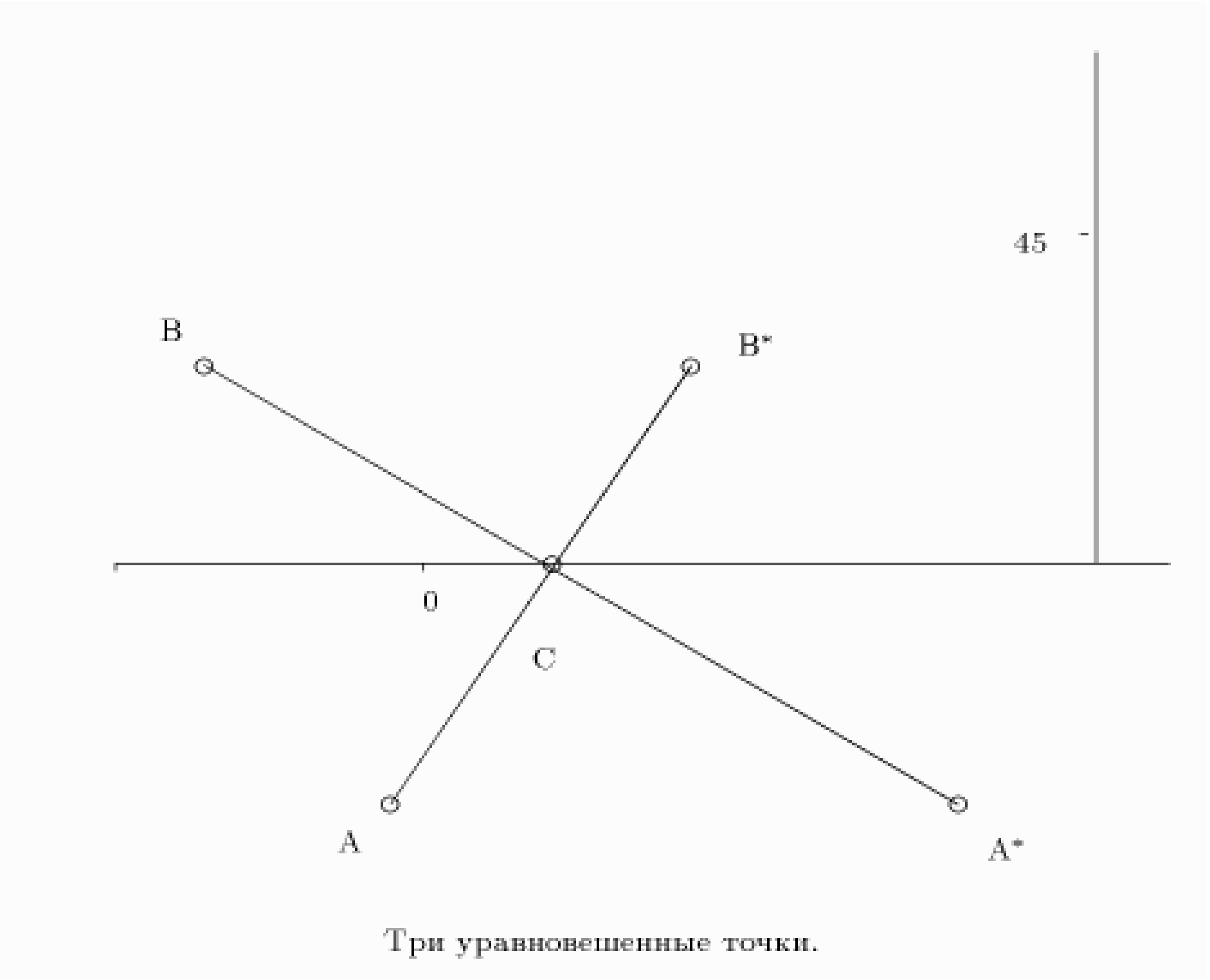

Рис. 34 Three balanced points

The solution for four given points, that are the vertices of the rectangle, is provided on figure 32.

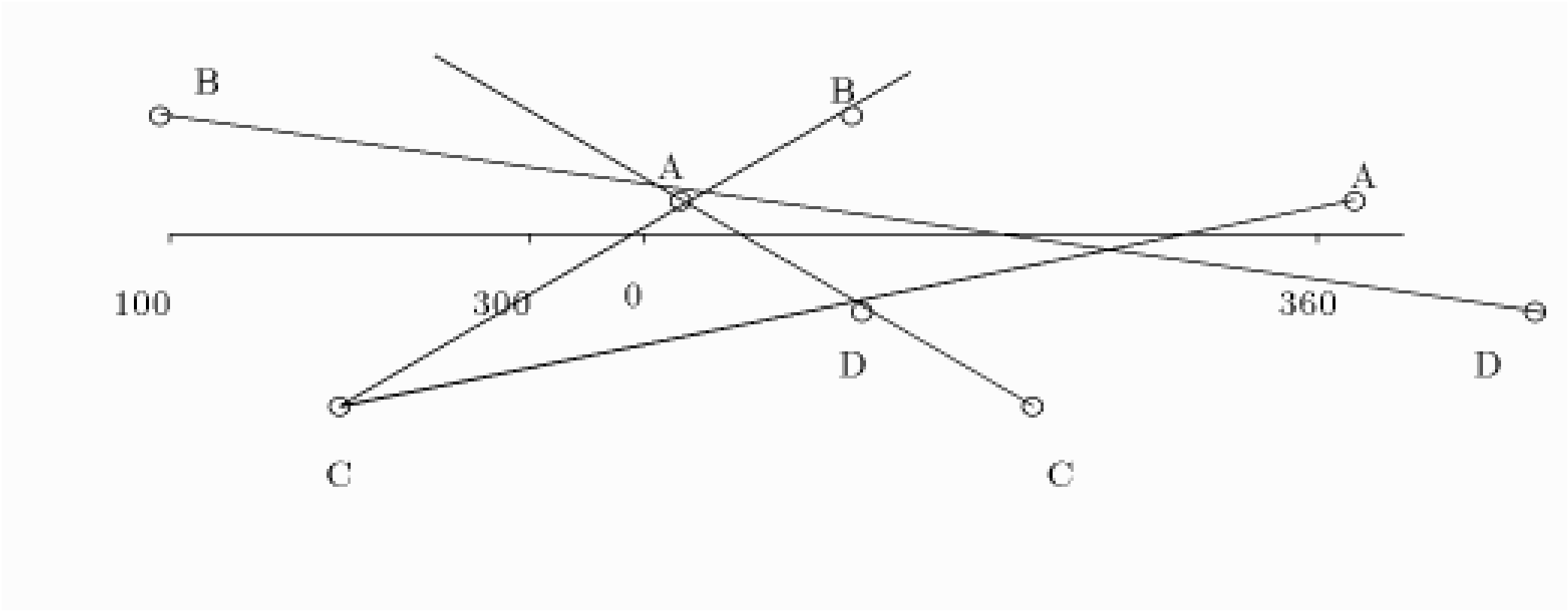

Fig. 35 Many balanced points

We are more interested in mechanic analogy, when the minimum of a distance is interpreted as some extrem energy principle for mechanic system. For example, Courant and Robbins (CR) [11] pointed out that if we consider Steiner's tree as nodes connected with an elastic thread, then the minimum of distance corresponds to the total potential energy of tension.

Considering the above mentioned, they constructed a device, that allows to find approximate solutions to the problem. CR used soap film, stretched between two pieces of glass, as a model, and interconnected with rods, which are located in places of given points. By lifting one piece of glass over the other, CR got some film that was interconnecting the rods. Because of experimental energy principle the position of the film provides one of the solutions of the problem. Unfortunately, it is not always possible to get the exact solution of the problem using this method, but we get something that is called Steiner minimal trees. It turns out that such trees can be obtained not only by using CR's device.

Imagine, that Steiner's tree has no vertices that have more then three line segments, so the additional point may only have 3 line segments that converge to it on 120° angle. Imagine, that the points of the tree are fixed and Steiner points "are looking for" static equilibrium, so the final configuration would satisfy the principle of potential gravitational energy ( the masses of points are considered equal and potential gravitational energy is negative).

Then we can easily demonstrate that any small departure from the configuration leads to the increase of total length of line segments of Steiner's tree. Moreover we can range Steiner minimal trees according to "force" that corresponds to given agitation.

Instead of calculating gravitational force, it is possible to determine the second derivative of coordinates by choosing for every Steiner point in iterative procedure a special locus with minimal total sum of second derivatives. Simply saying, by putting these points in the center of coordinates of the three closest neighbours. Then we can get an approximate solution of the problem. For example, the results of all previously studied cases, were achieved only after several (4-10) iterative tracing of variable points. It is obvious that three dimensional case (just like a case of higher dimension) doesn't not differ from two dimensional. By varying the initial approximations, we get several relatively minimal trees. If we get each of the trees more then one time, then the possibility of missing the most minimal tree is very little. This method may be applied to find other exact solutions for the Steiner problem (fig. 33)

Now, lets go back to the stars. Consider massive stars as "given" or fixed points of Steiner tree. We'll use smaller stars as additional Steiner points. Then the position with maximum potential energy will differ, because smaller stars or Steiner points will be placed in a specific way, when each of them will lie in the plane of its three closest neighbours.

So what does it imply? In spherical coordinate system, with the plane of one angle 1, coinciding with the plane formed by three neighbours, all the neighbouring stars will lie on a line 1 = 0. If we begin to turn the plane, so one of the stars will stay in the plane of angle 1, then, because of the symmetry, in a new spherical coordinate system, all three stars will lie on the same line, or, to be more specific, the stars will lie on two elements of a cylinder. It's the same as in the previously observed cases.

If we don't pay much attention to the condition that massive stars are motionless, but consider fixed only the location of the stars that lie on the boundary of some volume. In this case the symmetry in the spherical coordinate system also stays the same.

On figure 34, there is an example, when one point is balanced by three ones. On figure 35 there is an example, when out of 5 points we consider one of them as the location of viewer. On figure 36, there are the results of calculation for a large number of points.

As we can see, the figures, we got as the result of modeling, are very similar to the configurations, presented in the starry sky.

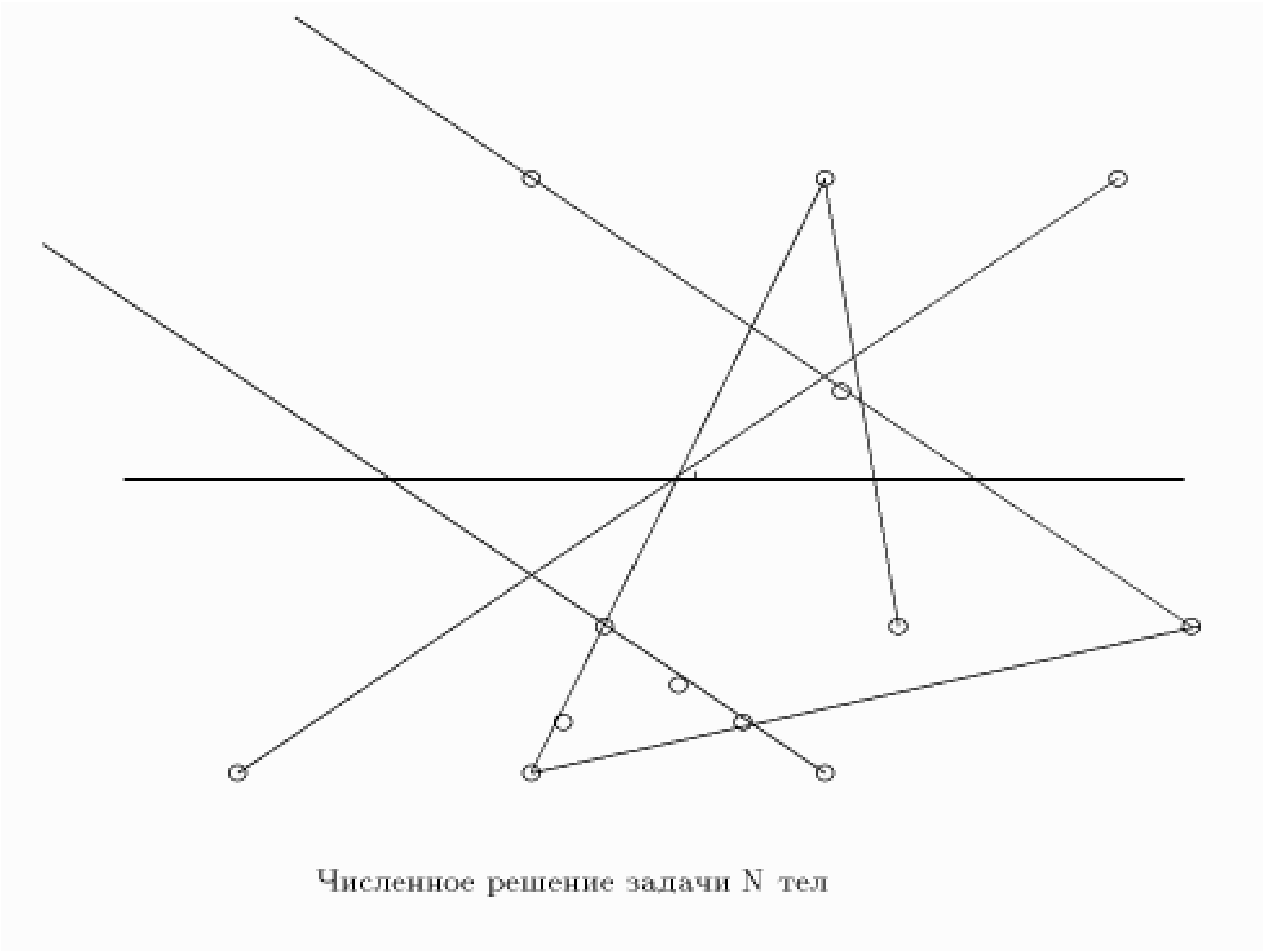

Fig. 36 numerical solution of the problem for N bodies**.**

The approbation of the above mentioned materials, where nothing was mentioned neither about SETI problem, nor about LDE, consisted not only in publications of the articles and presentations on the conferences, but also in some test to determine the importance of the results of the research, i.e. by observing the reaction of listeners and by considering their opinion on the level of the originality of this geometrical facts. Even more then publications, all this affected the value of achieved results for the author.

In other words it was a long term social-psychological experiment with a feedback.

We can state that the reaction of participants was mostly positive. Some was even saying: "It really is a discovery!" Even though there was no rejection of the stated facts, we still can't state that these facts gained interest of a broad spectrum of researchers, but it's not really important on this first stage.

A very important question still remains. Is it possible to get, an approximate, but some sort of information about the location of the planet or some the base of the sender of the probe? Certainly, here we enter a rather controversial sphere of interpretation of symbols, but we can still draw some conclusions.

Note that the major part of the delay time detected by the radio echo in 1920's in the experiments of Crawford equals 8. The star Procion is a binary star. One of its components a white dwarf, the other part is a star of spectral class F5 IV, which obviously have no planets suitable for living. But why do we conciser, that the signal should be transmitted from the native planet of a sender? From the point of view of the hypothesis a highly developed civilization does not really depend only on the resources of mother planet , this star can become a perfect temporary base, where any sputnik can be placed. Besides, next to a high-energy white dwarf the orbit of the sputnik will remain stable.

There definitely should be some clues to confirm that Procion is the right star. There are some. Lets consider the 5$^{th}$ series of Stormer, using which Lunen was able to make an assumption about the location of the senders. We'll try to prove that this series can contain some information about their location.

If we study the points we got when we draw lines by combining the points with coordinates that correspond to the numbers of stars, then we get triangles with a common vertex in a point that corresponds to the location of Procion.

Lets study different arrangement of stars according to their luminosity offered by Hipparchus (or the order determined by the photographic method of fixation of apparent luminosity). The projection of lines that connect the coordinates of stars on the plane of astronomical equator will form a sort of an "arrow", which goes through the center of the circumference and keeps its shape even when we build the same graph in ecliptic coordinate system and points to Procion. (Fig 37).

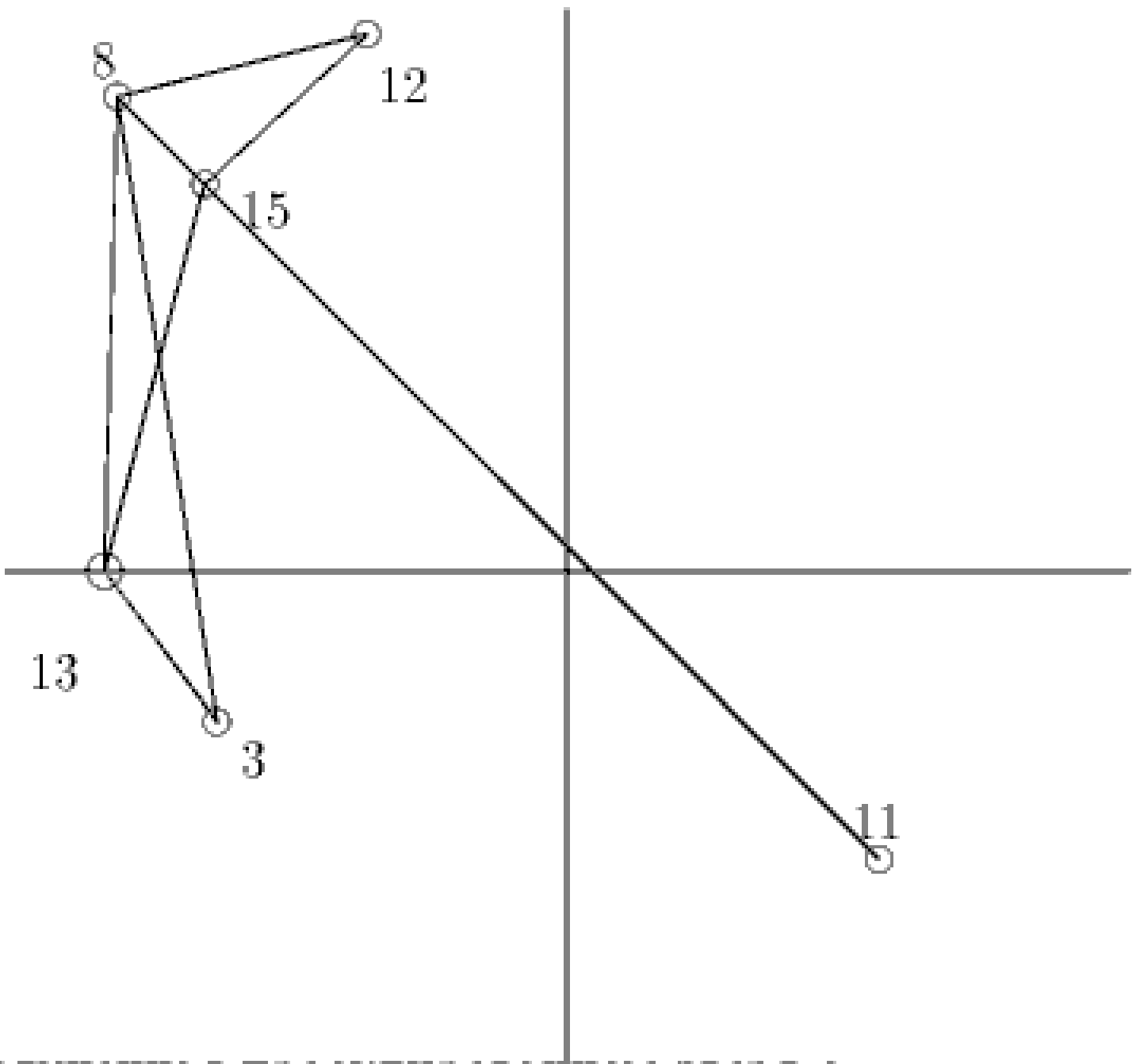

"Стрела" полученная проектированием звезд с номерами, отвечающими пятой серии Штермера, на плоскость экваториального круга 8-Процион 3 альфа Центравра, 11 Альтаир, 12 Бетельгейзе 13 альфа Южного Креста, 15 Поллукс (Гиппарх). "Стрела" сохраняется и на плоскости эклиптики.

Fig. 37 the “arrow” which was got by projecting stars with numbers that correspond to the 5th series of Stormer on the plane of the equatorial circle. 8-Procion, 3-Alfa Centauri, 11-Altair, 12-Betelgeuse, 13-Alfa Crux, 15-Pollux. The "arrow" keeps its shape in ecliptic coordinate system.

One may think that we depart from the requirement of objectivity of symbol, but is the symbol of an "arrow" so anthropomorphic? It is possible that the symbol of vector or directivity, as well as strictly determined mathematical figures, may also be universally accepted.

The other variant is also possible. Lets study the list of the closest stars and the 5th series:

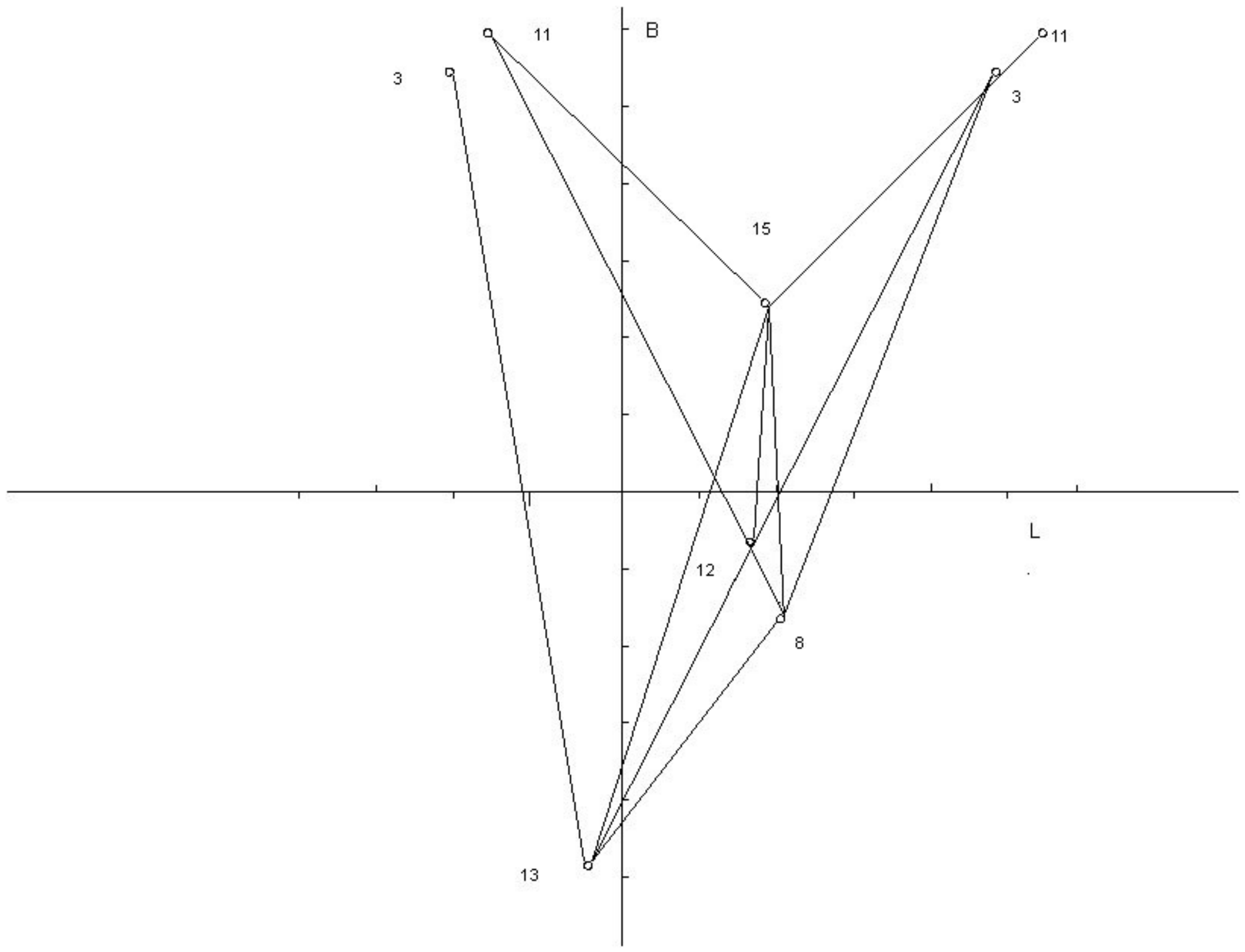

Fig. 38

Again we get a regular structure with 3 pairs "parallel segments" (Fig. 38) There is a point number 12 which lies on the intersection of the segments. This point corresponds to the stellar system of 61 Cygni, which has a habitable area. Note that the studies of the 5th series don't provide any regular structures for other lists.

The final step remains, does this all comes down only to stars? If there is a extraterrestrial automatic device, then it would be natural to expect more down to earth confirmation of the artificiality of the message. Indeed, when will the humanity be able to reach 61 Cygni or Procion?

What could we do if we were a hypothetic probe? We could choose some class of objects on the Moon, arrange them and place some artifact on the limited space of this object.

How can we choose this class of objects? We may consider some planetary ray structures on the Moon. There are not so many objects, and they are similar to the stars. Crater Tycho Brahe is the most famous of them. Lets arrange them according to the diameter of the crater and study them in the spherical coordinate system, but in connection with the Moon:

| Number | Name | l | b | Diameter |
|---|---|---|---|---|
| 1 | Langrenus | 241 | -9 | 140 |
| 2 | Tycho | 169 | -43 | 90 |
| 3 | Copernicus | 160 | 10 | 90 |
| 4 | Jackson | 1 | 22 | 85 |
| | | 7 | | |
| 5 | Ohm | 66 | 18 | 70 |
| 6 | Cavalerius | 113 | 5 | 65 |
| 7 | Anaxagoras | 170 | 73 | 55 |
| 8 | Aristillus | 181 | 34 | 55 |
| 9 | Olbers A. | 102 | 8 | 45 |
| 10 | Aristarchus | 132 | -24 | 40 |
| 11 | Thales | 230 | 62 | 35 |
| 12 | Kepler | 142 | 8 | 30 |
| 13 | Proclus | 227 | 16 | 30 |
| 14 | ------ | 72 | 44 | 30 |
| 15 | ------ | 302 | -3 | 25 |

The last figure presents a configuration that corresponds to the $5^{th}$ series of Stormer and demonstrates the properties of incidence and parallelism:

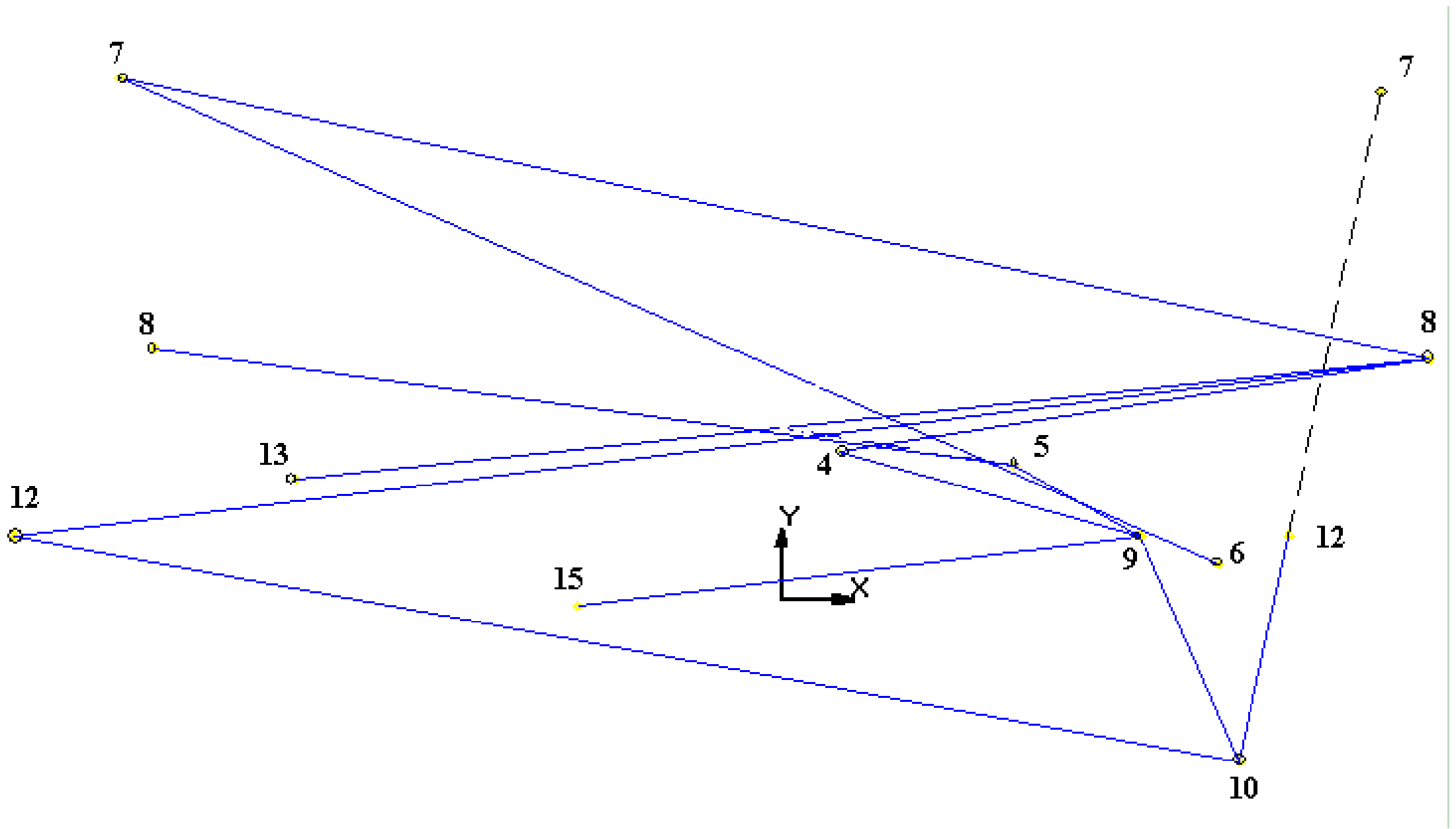

Note, that point number 8 corresponds to the Crater Aristillus, the first well-known Greek astronomer, who wrote his work "About fixed stars" even before Hipparchus. So we finally came back to stars again. If it really is a message, then it's done elegantly! Also, pay attention to the fact that this crater is only one degree shifted from the primary meridian (on the aim), so the location of the artefact may be either inside the crater or on the line of intersection of the primary meridian and the crater. This limits the area of search to several hundred square meters.

If we consider inverted sequence 15,9,4,8,13,8,12,10,9,5,8,7,6 → 5, 11, 16, 12, 7, 12, 10, 11, 15, 12, 13, 14 ( delay time complements to 20), then we get the following figure:

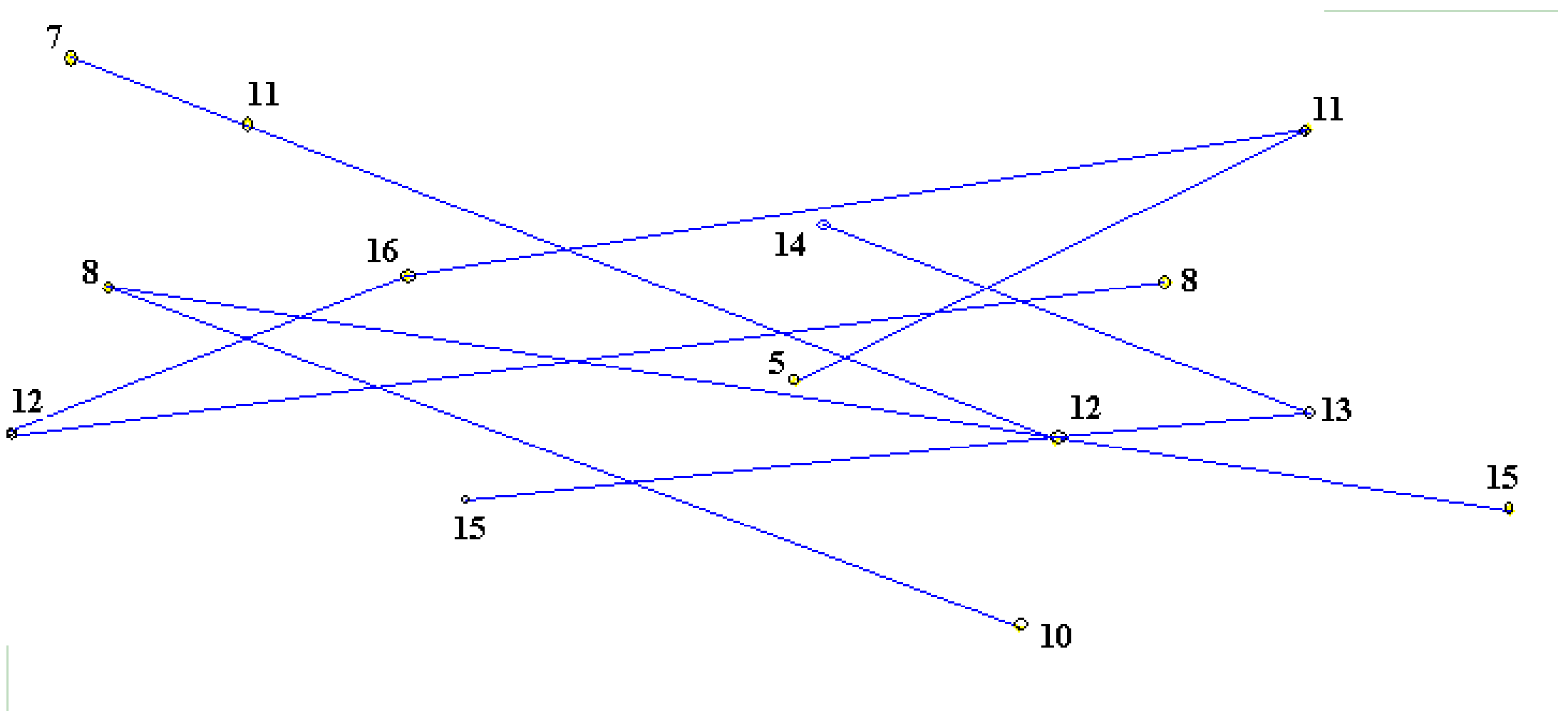

The second and the third series again illustrate the property of incidence, which means that points 12,5, 8 and12, 14, 8 are on the same line.

It is too early to claim, that presented material is indeed an interpretation of the message. But, in this work, we were able to prove that there is a connection between the numbers of delay time and the numbers of stars. This nontrivial results were developed in scientific publications.

We also believe that this way of studying of geometrical properties of figures in the given sphere is clearly connected with Gausse's idea (1826). He proposed to deforest an area in Siberian taiga in a shape of a right triangle, the sides of which correlate as 3, 4, 5.

The author expresses gratitude to M.Ja. Marov and L. M. Gindilis for  help questions  and support.

Project Meta's data. There was registered 37 strong strange 'alerts' similar to a famous signal WOW! (Horowitz&Sagan ApJ, 415, 218-235, 1993).

Five years of project Meta:an all-sky narrow-band radio search for extraterrestrial signals.

**1420 MHz**

00.82 3.25 7735

00.87 57.5 6942

06.08 -3.5 6782

06.23 9.5 6822

11.58 31.5 6876

21.15 -21.0 6737

21.98 38.5 6894

01.70 33.5 8014

02.90 32.0 8022

03.10 58.0 7847

12.32 16.0 8160

12.73 -12.5 8364

15.55 17.0 8154

19.57 47.5 7916

23.72 8.5 8216

**2840 MHz**

00.82 3.25 7735

01.30 -22.00 7577

01.83 7.0 7769

05.73 6.0 7326

08.00 -8.50 7415

08.03 11.00 7301

08.08 7.00 7769

08.67 45.75 7159

08.95 -15.75 7452

10.43 -21.25 7481

11.23 58.00 7230

14.30 57.50 7228

14.65 46.50 7164

15.47 -18.00 7599

17.10 2.00 7351

18.05 23.50 7061

18.37 -19.50 7467

18.45 38.50 7127

18.67 -23.25 7493

18.68 -22.25 7565

19.18 -0.50 7699

19.67 -23.00 7560

20.03 30.75 7092

01.30 -22.00 7577

01.83 7.0 7769

05.73 6.0 7326

08.00 -8.50 7415

08.03 11.00 7301

08.08 7.00 7769

08.67 45.75 7159

08.95 -15.75 7452

10.43 -21.25 7481

11.23 58.00 7230

14.30 57.50 7228

14.65 46.50 7164

15.47 -18.00 7599

17.10 2.00 7351

18.05 23.50 7061

18.37 -19.50 7467

18.45 38.50 7127

18.67 -23.25 7493

18.68 -22.25 7565

19.18 -0.50 7699

19.67 -23.00 7560

20.03 30.75 7092

**RA DEC Date**

00.87 57.5 6942

06.08 -3.5 6782

06.23 9.5 6822

11.58 31.5 6876

21.15 -21.0 6737

21.98 38.5 6894

01.70 33.5 8014

02.90 32.0 8022

03.10 58.0 7847

12.32 16.0 8160

12.73 -12.5 8364

15.55 17.0 8154

19.57 47.5 7916

23.72 8.5 8216

**2840 MHz**

00.82 3.25 7735

01.30 -22.00 7577

01.83 7.0 7769

05.73 6.0 7326

08.00 -8.50 7415

08.03 11.00 7301

08.08 7.00 7769

08.67 45.75 7159

08.95 -15.75 7452

10.43 -21.25 7481

11.23 58.00 7230

14.30 57.50 7228

14.65 46.50 7164

15.47 -18.00 7599

17.10 2.00 7351

18.05 23.50 7061

18.37 -19.50 7467

18.45 38.50 7127

18.67 -23.25 7493

18.68 -22.25 7565

19.18 -0.50 7699

19.67 -23.00 7560

20.03 30.75 7092